\documentclass[11pt,a4paper]{article}
\usepackage{jheppub,amsmath,amssymb,slashed,url,bm}
\usepackage{graphicx}
\usepackage{epstopdf}

\newcommand{\e}{\begin{eqnarray}}
\newcommand{\ee}{\end{eqnarray}}

\def\a{\alpha}
\def\b{\beta}
\def\d{\delta}

\newcommand{\g}{\gamma}

\newcommand{\s}{\sigma}

\newcommand{\vp}{\varepsilon}

\newcommand{\lds}{\Lambda_{\rm S}}
\newcommand{\ls}{L_{\rm S}}
\newcommand{\rs}{R_{\rm S}}

\newcommand{\ww}{\widetilde{W}}
\newcommand{\wl}{\widetilde{L}}

\newcommand{\hpl}{\widehat{p}_\Lambda}
\newcommand{\hp}{\widehat{p}}

\font\teneurm=eurm10 \font\seveneurm=eurm7  \font\fiveeurm=eurm5
\newfam\eurmfam
\textfont\eurmfam=\teneurm \scriptfont\eurmfam=\seveneurm
\scriptscriptfont\eurmfam=\fiveeurm

\font\teneusm=eusm10 \font\seveneusm=eusm7 \font\fiveeusm=eusm5
\newfam\eusmfam
\textfont\eusmfam=\teneusm \scriptfont\eusmfam=\seveneusm
\scriptscriptfont\eusmfam=\fiveeusm

\font\tencmmib=cmmib10 \skewchar\tencmmib='177
\font\sevencmmib=cmmib7 \skewchar\sevencmmib='177
\font\fivecmmib=cmmib5 \skewchar\fivecmmib='177
\newfam\cmmibfam
\textfont\cmmibfam=\tencmmib \scriptfont\cmmibfam=\sevencmmib
\scriptscriptfont\cmmibfam=\fivecmmib

\title{General Wigner Rotations in $D$ Dimensions}

 \author{Fa-Min Chen }
\affiliation{Department of Physics, Beijing Jiaotong University, Beijing 100044, China}
\abstract{
We construct general Wigner rotations for both massive and massless particles in $D$-dimensional spacetime. We work out the explicit expressions of these Wigner rotations for arbitrary  Lorentz transformations. We study the relation between the electromagnetic gauge invariance and the non-uniqueness of Wigner rotation.}
\begin{document} \maketitle

\section{Introduction and Summary}\label{secintro}
In quantum field theory, one-particle states are classified according to the representations of little groups of the Lorentz group \cite{Wigner}. For a systematic introduction of little groups or Wigner rotations for both massive and massless particles in four dimensional spacetime, see Ref. \cite{Weinberg1}.

In this paper, we wish to study the Wigner rotations for both massive and massless particles in an arbitrary $D$-dimensional spacetime. We begin by introducing the little groups in $D$-dimensional spacetime. For a chosen ``standard" $D$-momentum $k^\nu$ \footnote{For a particle of unit mass, $k^\mu=(0,0,\ldots,0,1)$; For a massless particle, $k^\mu=(0,\ldots,0,1,1)$, with ``1" standing for unit energy.}, the little group or Wigner rotation is defined as
$W^\mu{}_\nu k^\nu=k^\mu$, $\mu,\nu=0,1,\ldots, D-1.$
For an arbitrary Lorentz transformation $\Lambda$ and a given momentum $p^\mu$, the little group can be constructed as follows \cite{Weinberg1},
\e
W(\Lambda,p)=L^{-1}(\Lambda p)\Lambda L(p)\label{elements}.
\ee
Here $L(p)$ is some standard Lorentz transformation, bringing $k^\mu$ to $p^\mu$, i.e. $p^\mu=L^\mu{}_\nu(p)k^\nu$.

In this paper, we work out the explicit expressions of little group elements (\ref{elements}) for both massive and massless particles in $D$-dimensional spacetime.

Our main idea is to use spinor algebra to construct the little groups or Wigner rotations. Generally speaking, the spinor algebra in $D$ dimensions is slightly easier than the tensor algebra. Nevertheless, the spinors can still furnish faithful representations of the little groups; So they can be used to work out (\ref{elements}). The technical details will be introduced in the next section. For the massive particle case, we use two distinct methods to derive the explicit expression for the Wigner rotations; In the special case of $4D$, we provide a third way to work out the explicit expression for the Wigner rotation.


The spinor representation of little group for massless particles is particularly interesting. For instance, in the case of $4D$, the little group is $ISO(2)$, with the rotation generator $J^3$ and two translation generators $T^1$ and $T^2$. If the physical state is a superposition of the eigenvectors of $T^1$ and $T^2$, and if the eigenvalues of $T^1$ and $T^2$ are not zero, the helicity $\sigma$ of a massless particle would have a continuous value without taking account of the topology of the Lorentz group \cite{Weinberg1}. However, in the spinor realization of $ISO(2)$, the eigenvalues of $A^1\equiv T^1_{\rm S}$ and $A^2\equiv T^2_{\rm S}$ are zero automatically. (Here ``S" stands for the spinor representation.) So a continuous value of the helicity $\sigma$ of a massless fermionic particle can be avoided, without even considering the topology of the Lorentz group. 

It is obvious for a given Lorentz transformation, the Wigner rotation cannot be uniquely defined. For a fixed ``standard" $D$-momentum $k^\mu$, one may choose two different standard Lorentz transformations $L(p)$ and $\widetilde L(p)$, in the sense that $L(p)^\mu{}_\nu k^\nu=\widetilde L(p)^\mu{}_\nu k^\nu=p^\mu$ but $L(p)\neq\widetilde L(p)$. The resulting two Wigner rotations satisfy
\e
\widetilde{W}(\Lambda,p)=S(\Lambda p)W(\Lambda,p)S^{-1}(p)\label{2wigner1}
\ee
where $S(p)\equiv \widetilde L^{-1}(p)L(p)$. The above equation may be useful in studying gauge fields: Here $S(p)$ may have a connection with the gauge transformation of $U(1)$ gauge field in $D$ dimensions. As an example, we discuss the relation between the electromagnetic gauge invariance and the non-uniqueness of Wigner rotation in four dimensional spacetime (see Section \ref{Seclittgauge}).

The results of this paper may be useful in studying theories in the higher dimensions, such as superstring theory or M-theory.

Our paper is organized as follows. In Section \ref{Secmassive}, we work out the Wigner rotations for massive particles in $D$ dimensions, and discuss the special case of $D=4$. In Section \ref{Secmassless}, we derive the Wigner rotations for massless particles in $D$ dimensions; We investigate the special case of $D=4$, and study the relation between the Wigner rotation and the $U(1)$ gauge symmetry. We summarize our conventions and some useful identities in Appendix \ref{conventions}. In Appendix \ref{SecSODm1}, we verify that the little group elements for massive particles belong to  $SO(D-1)$, and  In Appendix \ref{SecSODm2}, we verify that some little group elements for massless particles belong to $SO(D-2)$.



\section{Wigner Rotations for Massive Particles}\label{Secmassive}
\subsection{$D$ Dimensions}
For a particle of mass $M$ in $D$ dimensions, we choose the standard vector as $k^\mu=(0,0,\ldots,0,M)$. The spinor representation of the ``standard boost" can be constructed as follows
\e
L_{\rm S}(\eta)=e^{\eta_i\Sigma^{i0}}\label{slrnt1}
\ee
Here $\Sigma^{i0}=\frac{1}{4}[\g^0,\g^i]$ is the set of boost generators (our conventions are summarized in Appendix \ref{conventions}), and $\eta^i$ the set of rapidities; The subscript ``S" stands for spinor representation. The relation between $L_{\rm S}(\eta)$  and $L(\eta)$ \footnote{In this paper, Lorentz transformations without the subscript ``S", such as $L(p)$, $\Lambda$, $R$, and $W(\Lambda,p)$ are in the vector representation.} is the standard one:
\e
L_{\rm S}(\eta)\g^\mu L^{-1}_{\rm S}(\eta)=L_\nu{}^\mu(\eta)\g^\nu.\label{spvc}
\ee
Using $(2\Sigma^{i0})^2=1$ (no sum), one can convert (\ref{slrnt1}) into the form
\e
L_{\rm S}(\eta)=\cosh(\eta/2)+\sinh(\eta/2)\hat\eta^i(2\Sigma^{i0}),\label{boost1}
\ee
where $\hat\eta^i\equiv \eta^i/\eta$ and $\eta\equiv |\vec\eta|=\sqrt{(\eta^i)^2}$. Substituting  (\ref{boost1}) into (\ref{spvc}), we find that
\e
&&L_i{}^j(\eta)=\d^{ij}+(\cosh\eta-1)\hat\eta^i\hat\eta^j\nonumber\\
&&L_0{}^i(\eta)=L_i{}^0(\eta)=-\hat\eta^i\sinh\eta\label{standardLM}\\
&&L_0{}^0(\eta)=\cosh\eta\nonumber
\ee
Substituting
\e
\hat\eta^i=\hat p^i,\quad\ \sinh\eta=|\vec p|/M\label{etap}
\ee
into (\ref{standardLM}),
\e
&&L_i{}^j(p)=\d^{ij}+(\g-1)\hat p^i\hat p^j,\nonumber\\
&&L_0{}^i(p)=L_i{}^0(p)=-\hat p^i\sqrt{\g^2-1}\label{standardLM2},\\
&&L_0{}^0(p)=\g,\nonumber
\ee
where $\g\equiv \sqrt{|\vec p|^2/M^2+1}=p^0/M$. We see that $L(\eta)$ or $L(p)$ does carry the $D$-momentum from $k^\mu$ to $p^\mu$. Since now, we do not distinguish $L(\eta)$ and $L(p)$. It can be seen that if $D=4$, the standard boost (\ref{standardLM}) is exactly the same as the one in Ref. \cite{Weinberg1}.

For a \emph{given} general Lorentz transformation $\Lambda$, we denote its spinor counterpart  as
$\Lambda_{\rm S}$;\footnote{If $D\leq 4$, it is relatively easy to  work out the explicit expression of $\Lambda_{\rm S}$ for a given general $\Lambda$. (See Section \ref{sect4D}.)} They satisfy the equation
\e
\lds\g^\mu\lds^{-1}=\Lambda_\nu{}^\mu\g^\nu\label{grntz1}
\ee
Then the Wigner rotation in the spinor space reads
\e
W_{\rm S}(\Lambda,\eta)=L^{-1}_{\rm S}(\eta_\Lambda)\Lambda_{\rm S}L_{\rm S}(\eta).\label{littlem}
\ee
Here $\eta_\Lambda$ must be defined such that $L(\eta_\Lambda)$ transforms $p^\mu$ into $(\Lambda p)^\mu$, i.e.,
\e
\hat\eta^i_\Lambda=\widehat{(\Lambda p)}^i,\quad\ \sqrt{ \big((\Lambda p)^i\big)^2}=M\sinh(\eta_\Lambda).
\ee
This can be fulfilled by requiring that
\e
L_{\rm S}(\eta_\Lambda)\g^0 L^{-1}_{\rm S}(\eta_\Lambda)=\Lambda_{\rm S}L_{\rm S}(\eta)\g^0 L^{-1}_{\rm S}(\eta)\Lambda^{-1}_{\rm S}
\ee
On one hand,
\e
\Lambda_{\rm S}L_{\rm S}(\eta)\g^0 L^{-1}_{\rm S}(\eta)\Lambda^{-1}_{\rm S}=\Lambda_\nu{}^\mu L_{\mu}{}^0(\eta)\g^\nu.\label{spvc2}
\ee
On the other hand,
in analogy to (\ref{boost1}), we have
\e
L_{\rm S}(\eta_\Lambda)&=&\cosh(\eta_\Lambda/2)+\sinh(\eta_\Lambda/2)\hat\eta^i_\Lambda(2\Sigma^{i0}).
\ee
So
\e
L_{\rm S}(\eta_\Lambda)\g^0 L^{-1}_{\rm S}(\eta_\Lambda)=L_\nu{}^0(\eta_\Lambda)\g^\nu&=&\bigg(\cosh(\eta_\Lambda)+\sinh(\eta_\Lambda)\hat\eta^i_\Lambda(2\Sigma^{i0})\bigg)\g^0\nonumber\\
&=&\cosh(\eta_\Lambda)\g^0-\sinh(\eta_\Lambda)\hat\eta^i_\Lambda\g^{i}.\label{spvc3}
\ee
Comparing (\ref{spvc2}) and (\ref{spvc3}) gives
\e
\cosh(\eta_\Lambda)&=&(\Lambda L)_0{}^0=\Lambda_0{}^0\cosh(\eta)-\Lambda_0{}^i\hat\eta_i\sinh(\eta),\nonumber\\
\hat\eta^j_\Lambda\sinh(\eta_\Lambda)&=&-(\Lambda L)_j{}^0=\Lambda_j{}^i\hat\eta_i\sinh(\eta)-\Lambda_j{}^0\cosh(\eta).\label{parameter1}
\ee
where we have used (\ref{standardLM}), and for readability, we have written $\Lambda_\mu{}^\rho L_\rho{}^\nu$ as $(\Lambda L)_\mu{}^\nu$.

The inverse transformation reads
\e
L^{-1}_{\rm S}(\eta_\Lambda)&=&\cosh(\eta_\Lambda/2)-\sinh(\eta_\Lambda/2)\hat\eta^i_\Lambda(2\Sigma^{i0}).\label{slrntz3}
\ee
It is possible to recast it into the following form:
\e
L^{-1}_{\rm S}(\eta_\Lambda)&=&\frac{(\Lambda^\dag_{\rm S})^{-1}L^{-2}_{\rm S}(\eta)\Lambda^\dag_{\rm S}+1}{2\cosh(\eta_\Lambda/2)}.\label{slrntz2}
\ee
To see this, let us evaluate $\Lambda_{\rm S}L^2_{\rm S}\lds^\dag$ first. Using (\ref{boost1}), (\ref{grntz1}), and $\lds^{-1}=\g^0\lds^\dag(\g^0)^{-1}$, we find that
\e
\Lambda_{\rm S}L^2_{\rm S}\lds^\dag&=&\Lambda_{\rm S}[\cosh(\eta)+\sinh(\eta)\hat\eta^i(2\Sigma^{i0})]\lds^\dag\\
&=&[\Lambda_0{}^0\cosh(\eta)-\Lambda_0{}^i\hat\eta^i\sinh(\eta)]+[\Lambda_j{}^i\hat\eta^i\sinh(\eta)
-\Lambda_j{}^0\cosh(\eta)](2\Sigma^{j0})\nonumber
\ee
Using the above result, it is not difficult to compute $(\Lambda^\dag_{\rm S})^{-1}L^{-2}_{\rm S}(\eta)\Lambda^\dag_{\rm S}$:
\e
&&(\Lambda^\dag_{\rm S})^{-1}L^{-2}_{\rm S}(\eta)\Lambda^\dag_{\rm S}\nonumber\\
&=&\bigg(\Lambda_{\rm S}L^2_{\rm S}\lds^\dag\bigg)^{-1}=\g^0\bigg(\Lambda_{\rm S}L^2_{\rm S}\lds^\dag\bigg)^\dag(\g^0)^{-1}\\
&=&[\Lambda_0{}^0\cosh(\eta)-\Lambda_0{}^i\hat\eta^i\sinh(\eta)]-[\Lambda_j{}^i\hat\eta^i\sinh(\eta)
-\Lambda_j{}^0\cosh(\eta)](2\Sigma^{j0})\nonumber
\ee
Plugging the above equation into (\ref{slrntz2}), and using (\ref{parameter1}), we find that (\ref{slrntz2}) is exactly the same as (\ref{slrntz3}).

Plugging (\ref{slrntz3}) into (\ref{littlem}) gives the spinor representation of the general Wigner rotation for massive particles in $D$ dimension:
\e
W_{\rm S}(\Lambda,\eta)=\frac{(\lds^\dag)^{-1}
\ls^{-1}(\eta)+\lds\ls(\eta)}{2\cosh(\eta_\Lambda/2)}=\frac{\g^0\lds
\ls(\eta)(\g^0)^{-1}+\lds\ls(\eta)}{\sqrt{2(1+[\Lambda L(p)]_0{}^0)}},\label{littlem1}
\ee
where we have written the denominator as
\e
2\cosh(\eta_\Lambda/2)=\sqrt{2(\cosh(\eta_\Lambda)+1)}=\sqrt{2(1+[\Lambda L(p)]_0{}^0)}.
\ee
It is easy to check that
\e
W_{\rm S}^\dag(\Lambda,\eta)=W_{\rm S}^{-1}(\Lambda,\eta)
\ee
So according to our convention in Appendix \ref{conventions}, $W_{\rm S}(\Lambda,\eta)$ must furnish a unitary representation of $SO(D-1)$.

The general Wigner rotation or the little group element $W(\Lambda,\eta)$ can be worked out via the equation:
\e
W_{\rm S}(\Lambda,\eta)\g^\mu W^{-1}_{\rm S}(\Lambda,\eta)=W_\nu{}^\mu(\Lambda,\eta)\g^\nu
\ee
First of all, if $\gamma^\mu=\g^0$, it is easy to verify that
\e
W_{\rm S}(\Lambda,\eta)\g^0W^{-1}_{\rm S}(\Lambda,\eta)=\g^0,
\ee
that is,
\e
W_0{}^0(\Lambda,\eta)=1,\quad\quad W_i{}^0(\Lambda,\eta)=0.
\ee
Secondly, if $\gamma^\mu=\g^i$, using (\ref{spvc}), (\ref{grntz1}), and the commutation relations in Appendix \ref{conventions}, we obtain
\e
W_{\rm S}(\Lambda,\eta)\g^i W^{-1}_{\rm S}(\Lambda,\eta)&=&W_\nu{}^i(\Lambda,\eta)\g^\nu=W_j{}^i(\Lambda,\eta)\g^j\nonumber\\
&=&\bigg(-\frac{[\Lambda L(\eta)]_0{}^i[\Lambda L(\eta)]_j{}^0}{1+[\Lambda L(\eta)]_0{}^0}+[\Lambda L(\eta)]_j{}^i\bigg)\g^j.
\ee

In summary,
\e
&&W_0{}^0(\Lambda,p)=1,\nonumber\\
&&W_i{}^0(\Lambda,p)=W_0{}^i(\Lambda,p)=0,\nonumber\\
&&W_j{}^i(\Lambda,p)=-\frac{[\Lambda L(p)]_0{}^i[\Lambda L(p)]_j{}^0}{1+[\Lambda L(p)]_0{}^0}+[\Lambda L(p)]_j{}^i.\label{gwm1}
\ee
We see that once the explicit expression for $\Lambda$ is known, one can calculate $W_j{}^i(\Lambda,p)$ immediately, without having to work out the explicit expression of $\lds$.

Using (\ref{etap}), a short calculation gives
\e
W_j{}^i(\Lambda,p)
&=&\frac{[-\Lambda_0{}^0p^i/M+\Lambda_0{}^i+(\g-1)\Lambda_0{}^k\hat p_k\hat p^i](\Lambda p)_j}{M+(\Lambda p)^0}\nonumber\\
&&-\Lambda_j{}^0p^i/M+(\g-1)\Lambda_j{}^k\hat p_k\hat p^i+\Lambda_j{}^i.\label{gwm2}
\ee

The Wigner rotation (\ref{gwm1}) can be also derived without relying on Clifford algebra. We begin by writing down the standard boost $L(\Lambda p)$:
\e
&&L^i{}_j(\Lambda p)=\d^{ij}+(\g_{\Lambda}-1)\widehat{\Lambda p}^i\widehat{\Lambda p}^j,\nonumber\\
&&L^0{}_i(\Lambda p)=L^i{}_0(\Lambda p)=\widehat{\Lambda p}^i\sqrt{\g_{\Lambda}^2-1}\label{standardLM3},\\
&&L^0{}_0(\Lambda p)=\g_{\Lambda},\nonumber
\ee
where
\e
&&\g_{\Lambda}=(\Lambda p)^0/M=[\Lambda L(p)]^0{}_0,\nonumber\\
&&\widehat{\Lambda p}^i=\frac{(\Lambda p)^i}{\sqrt{(\Lambda p)^j(\Lambda p)^j}}=\frac{[\Lambda L(p)]^i{}_0}{\sqrt{\g_{\Lambda}^2-1}}\label{plambda}
\ee
The inverse transformation $(L^{-1})^\mu{}_\nu(\Lambda p)$ are determined by the fundamental equation \begin{equation}
(L^{-1})^\mu{}_\nu(\Lambda p)=\eta^{\mu\rho}\eta_{\nu\s}L^\s{}_\rho(\Lambda p).
\end{equation}
Substituting (\ref{standardLM3}) into the above equation gives
\e
&&(L^{-1})^i{}_j(\Lambda p)=\d^{ij}+\frac{[\Lambda L(p)]^i{}_0[\Lambda L(p)]^j{}_0}{[\Lambda L(p)]^0{}_0+1},\nonumber\\
&&(L^{-1})^0{}_i(\Lambda p)=(L^{-1})^i{}_0(\Lambda p)=-[\Lambda L(p)]^i{}_0\label{standardLM4},\\
&&(L^{-1})^0{}_0(\Lambda p)=[\Lambda L(p)]^0{}_0,\nonumber
\ee
Substituting  (\ref{standardLM4}) into the equation
\e
W^\mu{}_\nu(\Lambda,p)=(L^{-1})^\mu{}_\rho(\Lambda p)\Lambda^\rho{}_\s L^\s{}_\nu(p),
\ee
after a slightly length algebra, one obtains
\e
&&W^0{}_0(\Lambda,p)=1,\nonumber\\
&&W^i{}_0(\Lambda,p)=W^0{}_i(\Lambda,p)=0,\nonumber\\
&&W^j{}_i(\Lambda,p)=-\frac{[\Lambda L(p)]^0{}_i[\Lambda L(p)]^j{}_0}{1+[\Lambda L(p)]^0{}_0}+[\Lambda L(p)]^j{}_i,\label{gwm3}
\ee
which are in agreement with (\ref{gwm1}).

Using $\Lambda^\mu{}_\rho\Lambda^\nu{}_\s\eta^{\rho\s}=\eta^{\mu\nu}$ and $L^\mu{}_\rho L^\nu{}_\s\eta^{\rho\s}=\eta^{\mu\nu}$, it is not difficult to verify that
\e
W^k{}_i(\Lambda,p)W^k{}_j(\Lambda,p)=\d_{ij}.\label{soDm1}
\ee
(For a  detailed proof, see Appendix \ref{SecSODm1}.) Namely, the little group is indeed $SO(D-1)$. Eqs. (\ref{soDm1}) are also consistent with the fact that $W^0{}_i(\Lambda,p)=0$ (see the second line of (\ref{gwm3})). Notice that
\e
\eta_{\mu\nu}W^\mu{}_i(\Lambda,p)W^\nu{}_j(\Lambda,p)=-W^0{}_i(\Lambda,p)W^0{}_j(\Lambda,p)+W^k{}_i(\Lambda,p)W^k{}_j(\Lambda,p)=\d_{ij},
\ee
which are exactly the same as Eqs (\ref{soDm1}) taking account of $W^0{}_i(\Lambda,p)=0$.

We now proceed to discuss two important special cases: $\Lambda$ is a general pure boost or a general pure rotation.

If $\lds$ is a pure rotation, i.e., $\lds=R_{\rm S}$, then by (\ref{reality2}), one has $(R^\dag_{\rm S})^{-1}=R_{\rm S}$. Plugging it into equation (\ref{littlem1}), and using (\ref{boost1}), we are led to
\e
W_{\rm S}(R,\eta)=\frac{R_{\rm S}[\ls^{-1}(\eta)+\ls(\eta)]}{2\cosh(\eta_\Lambda/2)}=\frac{\cosh(\eta/2)}{\cosh(\eta_\Lambda/2)}R_{\rm S}=R_{\rm S}.\label{littlemr}
\ee
In the last equity, $\cosh(\eta_\Lambda/2)=\cosh(\eta/2)$ can be proved as follows: If $\Lambda=R$, one has $\Lambda_0{}^0=1$ and $\Lambda_0{}^i=0$; Plugging them into the first equation of (\ref{parameter1}) proves $\cosh(\eta_\Lambda)=\cosh(\eta)$. Using (\ref{littlemr}), we find that
\e
W_{\rm S}(R,\eta)\g^iW^{-1}_{\rm S}(R,\eta)=W_\mu{}^i(R,\eta)\g^\mu=\rs\g^i\rs^{-1}=R_j{}^i\g^j
\ee
Namely, $W_0{}^i(R,\eta)=0$ and $W_j{}^i(R,\eta)=R_j{}^i$. That is
\e
W(R,\eta)=R.
\ee
(One can also prove the above equation by substituting $\Lambda=R$ into (\ref{gwm1}).) In other words, if $\Lambda$ is an arbitrary pure rotation $R$, the Wigner rotation $W(R,\eta)$ is exactly the same as $R$, independent of the parameter $\eta$ or momentum $p$. In $4D$, the above important equation is proved by using a different method \cite{Weinberg1}. We see that in $D$ dimensions,  this equation still holds.

However, we have to emphasize that $W(R,\eta)=R$ is due to the particular ``standard boost"(\ref{standardLM}) or (\ref{standardLM2}). If we use another ``standard boost"
$\widetilde{L}(p)$, satisfying $\widetilde L(p)^\mu{}_\nu k^\nu=L(p)^\mu{}_\nu k^\nu=p^\mu$, but $\widetilde L(p)\neq L(p)=(\ref{standardLM2})$, it is possible that $\widetilde{W}(R,\eta)\neq R$. This can be seen as follows: According to (\ref{2wigner1}),
\e
\widetilde{W}(R,p)=S(\Lambda p)W(R,p)S^{-1}(p)=S(\Lambda p)RS^{-1}(p)\label{2wigner2}
\ee
where $S(p)\equiv \widetilde L^{-1}(p)L(p)$; Generally speaking, $S(\Lambda p)RS^{-1}(p)\neq R$. 

If $\lds$ is a pure boost, i.e., $\lds=\ls(\xi)$, then by (\ref{reality2}), we have $\ls^\dag(\xi)=\ls(\xi)$ . Plugging this equation into (\ref{littlem1}),  we obtain
\e
W_{\rm S}(\xi,\eta)\equiv W_{\rm S}(\Lambda,\eta)\big|_{\Lambda=L(\xi)} =\frac{\ls^{-1}(\xi)
\ls^{-1}(\eta)+\ls(\xi)\ls(\eta)}{\sqrt{2(1+[\ls(\xi) L(\eta)]_0{}^0})}.
\ee
Using (\ref{boost1}) and (\ref{standardLM}), a short calculation gives
\e
W_{\rm S}(\xi,\eta)=\cos\bigg(\frac{\Theta}{2}\bigg)+\sin \bigg(\frac{\Theta}{2}\bigg)\frac{2\hat\xi_i\hat\eta_j\Sigma^{ij}}{\sqrt{1-(\hat\xi\cdot
\hat\eta)^2}}=\exp{\bigg(\Theta\frac{\hat\xi_i\hat\eta_j\Sigma^{ij}}{\sqrt{1-(\hat\xi\cdot
\hat\eta)^2}}\bigg)},\label{purerotation}
\ee
where $\Theta$ is defined via the equation
\e
\tan\bigg(\frac{\Theta}{2}\bigg)=\frac{\sinh(\xi/2)\sinh(\eta/2)\sqrt{1-(\hat\xi\cdot
\hat\eta)^2}}{\cosh(\xi/2)\cosh(\eta/2)+(\hat\xi\cdot
\hat\eta)\sinh(\xi/2)\sinh(\eta/2)}.\label{pureboost0}
\ee
Note that $W_{\rm S}(\xi,\eta)$ is invariant under the discrete transformation $\eta\rightarrow -\xi$ and $\xi\rightarrow\eta$, or $\eta\rightarrow \xi$ and $\xi\rightarrow-\eta$ (see (\ref{purerotation})), i.e.,
\e
W_{\rm S}(\xi,\eta)=W_{\rm S}(\eta,-\xi)=W_{\rm S}(-\eta,\xi).\label{parasymm}
\ee
Using
\e
W_{\rm S}(\xi,\eta)\g^iW^{-1}_{\rm S}(\xi,\eta)=W_j{}^i(\xi,\eta)\g^j\label{pureboost1}
\ee
and Eq. (\ref{tau}), we find that
\e
W_j{}^i(\xi,\eta)=\exp{\bigg(\Theta\frac{\hat\xi_k\hat\eta_l\tau^{kl}}{\sqrt{1-(\hat\xi\cdot
\hat\eta)^2}}\bigg)_j{}^i},\label{pureboost2}
\ee
where
\e
(\tau^{kl})_j{}^i=\d^{li}\d^k_j-\d^{ki}\d^l_j
\ee
 is the set of $SO(D-1)$ matrices, defined via Eq. (\ref{tau}). We see that $W_j{}^i(\xi,\eta)$ is a rotation on the $\eta$-$\xi$ plane, possessing the symmetry property $W_j{}^i(\xi,\eta)=W_j{}^i(\eta,-\xi)=W_j{}^i(-\eta,\xi)$.
  The explicit expression of $W_j{}^i(\xi,\eta)$ can be worked out by  either plugging (\ref{pureboost0}) into (\ref{pureboost1}), or expanding (\ref{pureboost2}) directly:
\e
 W_j{}^i(\xi,\eta)&=&\d^i_j+\sin\Theta\frac{\hat\xi_k\hat\eta_l}{\sqrt{1-(\hat\xi\cdot
\hat\eta)^2}}(\tau^{kl})_j{}^i+2(1-\cos\Theta)\frac{(\hat \xi_m\hat\eta_n)(\hat\xi_k\hat\eta_l)}{1-(\hat\xi\cdot
\hat\eta)^2}(\tau^{mn}\tau^{kl})_j{}^i\nonumber\\
&=&\d^i_j+\frac{(\cosh\eta-1)(\cosh\xi-1)[2(\hat\eta\cdot\hat\xi)\hat{\eta}^{(i}\hat{\xi}^{j)}-(
\hat\xi^i\hat\xi^j+\hat\eta^i\hat\eta^j)]}{1+\cosh\eta\cosh\xi+(\hat\eta\cdot\hat\xi)\sinh\eta\sinh\xi}
\nonumber\\
&&-\frac{2\hat{\eta}^{[i}\hat{\xi}^{j]}[\sinh\eta\sinh\xi+(\cosh\eta-1)(\cosh\xi-1)(\hat\eta\cdot\hat\xi)]}{1+\cosh\eta\cosh\xi+(\hat\eta\cdot\hat\xi)\sinh\eta\sinh\xi},
\label{pureboost3}
\ee
where $\hat{\eta}^{(i}\hat{\xi}^{j)}=(\hat\eta^i\hat\xi^j+\hat\eta^j\hat\xi^i)/2$ and $\hat{\eta}^{[i}\hat{\xi}^{j]}=(\hat\eta^i\hat\xi^j-\hat\eta^j\hat\xi^i)/2$. In deriving (\ref{pureboost3}), we have used (\ref{pureboost0}). The Wigner rotation (\ref{pureboost3}) can be also worked out by substituting the pure boost $\Lambda=L(\xi)$ into the general Wigner rotation (\ref{gwm1}).

\subsection{4 Dimensions}\label{sect4D}
For four dimensional spacetime, the Lorentz group $SO(3,1)=SU(2)\times SU(2)$. 
Since the irreducible representation of $SU(2)$ is well known, it is possible to work out the explicit expressions for
the irreducible unitary representations of any dimensionality of the little group $W(\Lambda, \eta)$ (see (\ref{4drep})).

Our first goal is to work out the explicit expression of the spinor representation of the little group (\ref{littlem1}). We begin by calculating the general Lorentz transformation in spinor space
\e
\Lambda_{\rm S}=\exp\bigg({\frac{1}{2}\omega_{\mu\nu}\Sigma^{\mu\nu}}\bigg).\label{4dltr}
\ee
To simplify calculations, we decompose the generators $\Sigma^{\mu\nu}$ and the parameters into the irreducible parts,
\e
\Sigma^{\mu\nu}_{\pm}&=&\frac{1}{2}\bigg(\Sigma^{\mu\nu}\pm\frac{i}{2}\vp^{\mu\nu\rho\s}\Sigma_{\rho\s}\bigg),\\
\omega^{\mu\nu}_{\pm}&=&\frac{1}{2}\bigg(\omega^{\mu\nu}\pm\frac{i}{2}\vp^{\mu\nu\rho\s}\omega_{\rho\s}\bigg),\label{omgpm}
\ee
where the totally antisymmetric tensor is defined as $\vp^{0123}=-\vp_{0123}=1$. Notice that they satisfy the duality conditions
\e
\Sigma^{\mu\nu}_{\pm}&=&\pm\frac{i}{2}\vp^{\mu\nu\rho\s}\Sigma_{\pm\rho\s},\\
\omega^{\mu\nu}_{\pm}&=&\pm\frac{i}{2}\vp^{\mu\nu\rho\s}\omega_{\pm\rho\s}.\label{dualomg}
\ee
 Now the general Lorentz transformation (\ref{4dltr}) reads
\e
\Lambda_{\rm S}=\exp\bigg({\frac{1}{2}\omega_{+\mu\nu}\Sigma^{\mu\nu}_+}\bigg)\exp\bigg({\frac{1}{2}\omega_{-\mu\nu}\Sigma^{\mu\nu}_-}\bigg)
\ee
Define
\e
\Lambda_{\rm S\pm}&\equiv&\exp\bigg({\frac{1}{2}\omega_{\pm\mu\nu}\Sigma^{\mu\nu}_\pm}\bigg)
\ee
and
\e
\omega_\pm=\sqrt{\omega_{\pm\mu\nu}\omega^{\mu\nu}_\pm}\quad{\rm and}\quad \hat\omega_{\pm\mu\nu}=\frac{\omega_{\pm\mu\nu}}{\omega_\pm}.\label{homega}
\ee
Note that $\Lambda_{\rm S\pm}$ are nothing but the $SL(2,C)$ matrices. By a direct calculation, we find that
\e
\Lambda_{\rm S\pm}=\frac{1}{2}(1\mp i\g_5)\cos\frac{\omega_\pm}{2}+\hat\omega_{\pm\mu\nu}\Sigma^{\mu\nu}_\pm\sin\frac{\omega_\pm}{2}
+\frac{1}{2}(1\pm i\g_5),
\ee
where $\g_5$ is defined as $\g_5\equiv\g_0\g_1\g_2\g_3$, or $\g_5=\frac{1}{4!}\vp_{\mu\nu\rho\s}\g^{\mu}\g^\nu\g^\rho\g^{\s}$.
Using the above equations, it is not difficult to work out $\Lambda_{\rm S}$,
\e
\Lambda_{\rm S}&=&\Lambda_{\rm S+}\Lambda_{\rm S-}\nonumber\\
&=&\frac{1}{2}(1- i\g_5)\cos\frac{\omega_+}{2}+\frac{1}{2}(1+ i\g_5)\cos\frac{\omega_-}{2}\nonumber\\
&&+\hat\omega_{+\mu\nu}\Sigma^{\mu\nu}_+\sin\frac{\omega_+}{2}
+\hat\omega_{-\mu\nu}\Sigma^{\mu\nu}_-\sin\frac{\omega_-}{2}.\label{4dstr}
\ee
The vector counterparts of  $\Lambda_{\rm S\pm}$ are defined via the equations
\e
\Lambda_{\rm S\pm}\g^\nu \Lambda^{-1}_{\rm S\pm}\equiv\Lambda_{\pm\mu}{}^\nu\g^\mu.
\ee
A straightforward computation gives
\e
\Lambda_{\pm\mu}{}^\nu=\cos\bigg(\frac{\omega_{\pm}}{2}\bigg)
\d_\mu{}^\nu+2\sin\bigg(\frac{\omega_{\pm}}{2}\bigg)\hat\omega_{\pm\mu}{}^\nu.\label{so3p}
\ee
That is,
\e
&&\Lambda_{\pm0}{}^0=\cos(\omega_{\pm}/2)\nonumber\\
&&\Lambda_{\pm0}{}^i=\Lambda_{\pm i}{}^0=-2\sin(\omega_{\pm}/2)\hat\omega_{\pm i0}\nonumber\\
&&\Lambda_{\pm i}{}^j=\cos(\omega_{\pm}/2)\d^{ij}\mp 2i\sin(\omega_{\pm}/2)\vp^{ijk}\hat\omega_{\pm k0}\label{lambdapm}
\ee
If it is a pure boost, i.e., $\omega_{ij}=0$ and $\omega_{i0}\rightarrow\eta^i$, one has
\e
&&L_{\pm0}{}^0(\eta)=\cosh(\eta/2)\nonumber\\
&&L_{\pm0}{}^i(\eta)=L_{\pm i}{}^0(\eta)=-\sinh(\eta/2)\hat\eta^i\nonumber\\
&&L_{\pm i}{}^j(\eta)=\cosh(\eta/2)\d^{ij}\mp i\sinh(\eta/2)\vp^{ijk}\hat\eta^k\label{lpm}
\ee
The standard Lorentz transformation (\ref{standardLM}) can be also derived using $L_\mu{}^\nu(\eta)=\big(L_{-}L_{+}\big)_\mu{}^\nu(\eta)$ and (\ref{lpm}).
The vector counterpart of  $\Lambda_{\rm S}$ is given by
\e
\Lambda_{\mu}{}^\nu&=&\big(\Lambda_{-}\Lambda_{+}\big)_\mu{}^\nu\nonumber\\
&=&\cos\bigg(\frac{\omega_+}{2}\bigg)\cos\bigg(\frac{\omega_-}{2}\bigg)\d_\mu{}^\nu
+2\cos\bigg(\frac{\omega_+}{2}\bigg)\sin\bigg(\frac{\omega_-}{2}\bigg)\hat\omega_{-\mu}{}^\nu\nonumber\\
&&+2\cos\bigg(\frac{\omega_-}{2}\bigg)\sin\bigg(\frac{\omega_+}{2}\bigg)\hat\omega_{+\mu}{}^\nu
+4\sin\bigg(\frac{\omega_+}{2}\bigg)\sin\bigg(\frac{\omega_-}{2}\bigg)(\hat\omega_+\hat\omega_{-})_\mu{}^\nu\label{lambdavc}
\ee
Alternatively, using the relation between $SL(2,C)$ and the 4D Lorentz group, one can calculate $\Lambda_\mu{}^\nu$ as follows,
\e
\Lambda_{S\pm}(\overline\gamma_{\pm}\gamma^\nu)\Lambda^{-1}_{S\mp}=\Lambda_\mu{}^\nu(\overline\gamma_{\pm}\gamma^\mu),
\ee
where
\e
\overline\gamma_\pm\equiv\frac{1}{2}(1\mp i\gamma_5).
\ee

We now would like to work out the spinor little group (\ref{littlem1}). We expect that it takes the ``standard" form
\e
W_{\rm S}(\Lambda,\eta)=\frac{\g^0\lds
\ls(\eta)(\g^0)^{-1}+\lds\ls(\eta)}{\sqrt{2(1+[\Lambda L(p)]_0{}^0)}}=\cos\frac{\Theta}{2}+\sin\frac{\Theta}{2}\hat\Theta_i(2\Sigma_i)\label{standardform}
\ee
where
\e
\Sigma_i\equiv\frac{1}{2}\vp_{ijk}\Sigma^{jk},\quad \Theta_i\equiv\frac{1}{2}\vp_{ijk}\Theta^{jk},\quad
\hat\Theta_i\equiv\Theta_i/\sqrt{\Theta^2},
\ee
and $\Theta^i=\Theta^i(\Lambda,\eta)$ is a function of $\Lambda_{\mu}{}^\nu$ and $\eta^i$. To determine $\Theta^i$,
let's first calculate $\lds\ls(\eta)$. According to (\ref{4dstr}), it must take the general form
\e
\lds\ls(\eta)&=&\frac{1}{2}(1- i\g_5)\cos\frac{\a_+}{2}+\frac{1}{2}(1+ i\g_5)\cos\frac{\a_-}{2}\nonumber\\
&&+\hat\a_{+\mu\nu}\Sigma^{\mu\nu}_+\sin\frac{\a_+}{2}
+\hat\a_{-\mu\nu}\Sigma^{\mu\nu}_-\sin\frac{\a_-}{2}.\label{lmdlh}
\ee
where the new parameters $\hat\a_{\pm\mu\nu}=\hat\a_{\pm\mu\nu}(\omega,\eta)$ and $\a_{\pm}=\a_{\pm}(\omega,\eta)$ are functions of $\omega_{\mu\nu}$ and $\eta_i$, to be determined later. The definitions and properties of $\hat\a_{\pm\mu\nu}$ and $\a_{\pm}$ are similar to that of $\hat\omega_{\pm\mu\nu}$ and $\omega_{\pm}$ (see (\ref{omgpm}), (\ref{dualomg}), and (\ref{homega})). 
Inserting (\ref{lmdlh}) into the first equation of (\ref{standardform}),
\e
W_{\rm S}(\Lambda,\eta)=\frac{(\cos\frac{\a_+}{2}+\cos\frac{\a_-}{2})
-2i(\sin\frac{\a_+}{2}\hat\a_{+i0}+\sin\frac{\a_-}{2}\hat\a_{+i0})(2\Sigma_i)}{\sqrt{2(1+[\Lambda L(p)]_0{}^0)}}\label{wigners}
\ee

We now must determine the relations of $\a_{\pm\mu\nu}$ between $\omega_{\pm\mu\nu}$ and $\eta_{i}$.
According to (\ref{so3p}), the vector representations of $\Lambda_{\pm}L_{\pm}(\eta)$ are given by
\e
(\Lambda_{\pm}L_{\pm}(\eta))_{\mu}{}^\nu=\cos\bigg(\frac{\a_{\pm}}{2}\bigg)
\d_\mu{}^\nu+2\sin\bigg(\frac{\a_{\pm}}{2}\bigg)\hat\a_{\pm\mu}{}^\nu.\label{vecrep1}
\ee
Substituting (\ref{lambdapm}) and (\ref{lpm}) into (\ref{vecrep1}), one obtains
\e
(\Lambda_{\pm}L_{\pm}(\eta))_{0}{}^0=\cos\frac{\a_\pm}{2}=\cos\frac{\omega_\pm}{2}
\cosh\frac{\eta}{2}+2\sin\frac{\omega_\pm}{2}
\sinh\frac{\eta}{2}(\hat\omega_{\pm}\cdot\hat\eta)\label{cosa}
\ee
and
\e
(\Lambda_{\pm}L_{\pm}(\eta))_{0}{}^i
&=&-2\sin\frac{\a_\pm}{2}\hat\a_{\pm i0}\label{sina}\\
&=&-\cos\frac{\omega_\pm}{2}
\sinh\frac{\eta}{2}\eta_i-2\sin\frac{\omega_\pm}{2}
\cosh\frac{\eta}{2}\hat\omega_{\pm i0}\mp 2i\sin\frac{\omega_\pm}{2}
\sinh\frac{\eta}{2}(\hat\omega_{\pm}\times\hat\eta)_i,\nonumber
\ee
where $\hat\omega_{\pm}\cdot\hat\eta\equiv \hat\omega_{\pm i0}\hat\eta_i$ and $(\hat\omega_{\pm}\times\hat\eta)_i=\vp_{ijk}\omega_{\pm j0}\hat\eta_k$. Using the above two equations, all terms in the numerator of (\ref{wigners}) can be expressed in terms of $\omega_{\pm\mu\nu}$ and $\eta_{i}$.

Using (\ref{cosa}) and (\ref{sina}), we see that (\ref{wigners}) also takes the following form:
\begin{equation}\label{wigners2}
W_{\rm S}(\Lambda,\eta)=\frac{(\Lambda_{+}L_{+}(\eta))_{0}{}^0+(\Lambda_{-}L_{-}(\eta))_{0}{}^0}{\sqrt{2(1+[\Lambda L(p)]_0{}^0)}}+i\frac{
(\Lambda_{+}L_{+}(\eta))_{0}{}^i+(\Lambda_{-}L_{-}(\eta))_{0}{}^i}{\sqrt{2(1+[\Lambda L(p)]_0{}^0)}}(2\Sigma_i).
\end{equation}
Here
\e
\sqrt{2(1+(\Lambda L)_0{}^0)}=\sqrt{2[1+\Lambda_0{}^0\cosh(\eta)-\Lambda_0{}^i\hat\eta_i\sinh(\eta)]}\label{deno}
\ee
(See the first equation of (\ref{parameter1})).

Using (\ref{cosa}), (\ref{sina}), and (\ref{deno}), Eq. (\ref{standardform}) or (\ref{wigners2}) can be readily worked out:
\e
W_{\rm S}(\Lambda,\eta)=\cos\frac{\Theta}{2}+\sin\frac{\Theta}{2}\hat\Theta_i(2\Sigma_i)=\exp({\Theta\hat\Theta_i\Sigma_i}),
\label{3rdway}
\ee
where
\e
\cos\frac{\Theta}{2}=\frac{[(\cos\frac{\omega_+}{2}+\cos\frac{\omega_-}{2})\cosh\frac{\eta}{2}+
2(\sin\frac{\omega_+}{2}(\hat\omega_+\cdot\hat\eta)+\sin\frac{\omega_-}{2}(\hat\omega_-\cdot\hat\eta))\sinh\frac{\eta}{2}]}{\sqrt{2[1+\Lambda_0{}^0\cosh(\eta)-\Lambda_0{}^i\hat\eta_i\sinh(\eta)]}},
\label{cosah}
\ee
and
\e
\label{sinah}
&&\sin\frac{\Theta}{2}\hat\Theta_i\nonumber\\
&=&\frac{1}{\sqrt{2[1+\Lambda_0{}^0\cosh(\eta)-\Lambda_0{}^i\hat\eta_i\sinh(\eta)]}}\bigg[-i\bigg((\cos\frac{\omega_+}{2}-\cos\frac{\omega_-}{2})\sinh\frac{\eta}{2}\hat\eta_i\nonumber\\
&&+
2(\sin\frac{\omega_+}{2}\hat\omega_{+i0}+\sin\frac{\omega_-}{2}\hat\omega_{-i0})\cosh\frac{\eta}{2}+
2i(\sin\frac{\omega_+}{2}\hat\omega_{+j0}+\sin\frac{\omega_-}{2}\hat\omega_{-j0})\vp_{ijk}\eta_k\sinh\frac{\eta}{2}\bigg)\bigg].\nonumber\\
\ee
In (\ref{cosah}) and (\ref{sinah}), the set of parameters $\eta^i$ is related to the momentum $\vec p$ and mass $M$ via (\ref{etap}), and the relation between $\Lambda_\mu{}^\nu$ and $\omega_{\mu\nu}$ is given by (\ref{lambdavc}).

Note that Eq. (\ref{3rdway}) provides a third way to construct the vector representation of the little group (\ref{gwm1}) in four dimensional spacetime. Since now $\cos\frac{\Theta}{2}$ and $\sin\frac{\Theta}{2}\hat\Theta_i$ have been worked out completely, it is not difficult to complete the calculation
\e
W_j{}^i(\Lambda,\eta)=\bigg(\exp({\Theta\hat\Theta_k\tau^k})\bigg)_j{}^i
=\cos\Theta\d_{ji}+(1-\cos\Theta)\hat\Theta_j\hat\Theta_i+\sin\Theta\vp_{jik}\Theta^k,\label{4dvm}
\ee
where $\tau^k=\frac{1}{2}\vp^{kij}\tau_{ij}$, with $(\tau_{ij})_{kl}=\d_{ik}\d_{jl}-\d_{jk}\d_{il}$. Plugging the data of (\ref{cosah}) and (\ref{sinah}) into (\ref{4dvm}), a slightly length calculation gives
\e
W_j{}^i(\Lambda,\eta)=-\frac{[\Lambda L(\eta)]_0{}^i[\Lambda L(\eta)]_j{}^0}{1+[\Lambda L(\eta)]_0{}^0}+[\Lambda L(\eta)]_j{}^i,
\ee
which is exactly the same as (\ref{gwm1}) or (\ref{gwm2}), with $i,j=1, 2, 3$.

The Wigner rotation in any irreducible representation can be constructed by replacing $\tau_i\rightarrow -iJ_i$ in the right-hand side of the first equity of (\ref{4dvm}):
 \e
W^{(j)}_{m'm}(\Lambda,\eta)\equiv W^{(j)}_{m'm}\bigg(\Theta(\Lambda,\eta)\bigg)=\bigg(\exp(-i\Theta\hat\Theta_kJ^{(j)}_k)\bigg)_{m'm}.\label{4danyrep}
\ee
 Here the irreducible representations of $J_i$ are the familiar ones,
\e
(J_3^{(j)})_{m'm}
= m\hbar\d_{m'm},\quad(J_1^{(j)}\pm iJ_2^{(j)})_{m'm}
=\hbar\d_{m',m\pm1}\sqrt{(j\pm m+1)(j\mp m)},
\ee
where $m', m=j, j-1,\ldots,-(j-1),-j$.
The Wigner's formula for d-function may be useful in calculating $W^{(j)}_{m'm}(\Lambda,\eta)$. For instance, in the special case of $\hat\Theta_k=\hat y$ or $\Theta\hat\Theta_kJ^{(j)}_k=\Theta J^{(j)}_2$, Eq. (\ref{4danyrep}) is nothing but the
the Wigner's d-function \cite{Wigner2}:
 \e
W^{(j)}_{m'm}\bigg(\Theta(\Lambda,\eta)\bigg)&=&\sum_{k}(-1)^{k-m+m'}\frac{\sqrt{(j+m)!(j-m)!(j+m')!(j-m')!}}
      {(j+m-k)!k!(j-k-m')!(j-m+m')!}\nonumber\\
      &&\times \bigg(\cos\frac{\Theta}{2}\bigg)^{2j-2k+m-m'}\bigg(\sin\frac{\Theta}{2}\bigg)^{2k-m+m'},\label{4drep}
 \ee
where the expressions of $\cos\frac{\Theta}{2}$ and $\sin\frac{\Theta}{2}$ are given by (\ref{cosah}) and (\ref{sinah}).

\subsection{Summary of This Section}
In summary, in $D$ dimensions, the spinor representation of the Wigner rotation is given by
\e
W_{\rm S}(\Lambda,\eta)=\frac{\g^0\lds
\ls(\eta)(\g^0)^{-1}+\lds\ls(\eta)}{\sqrt{2(1+[\Lambda L(p)]_0{}^0)}},
\ee
and the vector representation of the Wigner rotation is given by
\e
W_j{}^i(\Lambda,p)&=&-\frac{[\Lambda L(p)]_0{}^i[\Lambda L(p)]_j{}^0}{1+[\Lambda L(p)]_0{}^0}+[\Lambda L(p)]_j{}^i\nonumber\\
&=&\frac{[-\Lambda_0{}^0p^i/M+\Lambda_0{}^i+(\g-1)\Lambda_0{}^k\hat p_k\hat p^i](\Lambda p)_j}{M+(\Lambda p)^0}\nonumber\\
&&-\Lambda_j{}^0p^i/M+(\g-1)\Lambda_j{}^k\hat p_k\hat p^i+\Lambda_j{}^i.
\ee
Here $\Lambda_\mu{}^\nu$ is an arbitrary Lorentz transformation, and $L(p)$ or $L(\eta)$
carries the standard $D$-momentum $k^\mu=(0,0,\ldots,0,M)$ to $p^\mu$, i.e. $L^\mu{}_\nu(\eta)k^\nu=p^\mu$, with $p^\mu$ the $D$-momentum of the particle of mass $M$. The explicit expressions of $L(p)$ and $L(\eta)$ are given by (\ref{standardLM})$-$(\ref{standardLM2}).
And $\lds$ and $
\ls(\eta)$ are spinor counterparts of $\Lambda_{\mu}{}^\nu$ and $L_\mu{}^\nu(\eta)$, respectively. The explicit expression for $\ls(\eta)$ is given by (\ref{boost1}).

\section{Wigner Rotations for Massless Particles}\label{Secmassless}
\subsection{$D$ Dimensions}
We now turn to the case of massless particles in $D$-dimensions. We define the standard $D$-vector of energy $\kappa$ as
\e
k^\mu=(0,0,\ldots, \kappa, \kappa).
\ee
We see that $k_\mu\g^\mu=\kappa(-\g^0+\g^{D-1})$.
It is therefore more convenient to work in the light-cone coordinates (our conventions are summarized in Appendix \ref{conventions}),
\e
\g^{\pm}=\frac{1}{\sqrt 2}(\pm\g^0+\g^{D-1}),\quad k^{\pm}=\frac{1}{\sqrt 2}(k^0+k^{D-1}).
\ee
In the light-cone coordinates, we have
\e
k_\mu\g^\mu=k_-\g^-=\sqrt{2}\kappa\g^-.
\ee
The little group $W^\mu{}_\nu$ preserves $k^\mu$, in the sense that $W^\mu{}_\nu k^\nu=k^\mu$. In the spinor space, this is equivalent to require that
\e
W_{\rm S}\g^-W_{\rm S}^{-1}=\g^-,\label{lgml}
\ee
where $W_{\rm S}$ is spinor representation of the little group.

We define the ``standard Lorentz transformation" in spinor space as follows
\e
L_{\rm S}(\lambda)&\equiv&\exp(\lambda_a\Sigma^{+a})\exp(\lambda_{-}\Sigma^{+-}),\label{ssf1}\\
&=&\cosh\frac{\lambda_-}{2}+e^{-\lambda_-/2}\lambda_a\Sigma^{+a}+2\sinh\frac{\lambda_-}{2}\Sigma^{+-}\nonumber.
\ee
where the set of generators is $(\Sigma^{+a},\Sigma^{+-})$, $a=1,\ldots,D-2$, with $\Sigma^{+a}=\frac{1}{\sqrt2}(\Sigma^{0a}+\Sigma^{D-1,a})$ and $\Sigma^{+-}=\Sigma^{0,D-1}$, and the parameters are defined as \footnote{For a massless particle of unit energy, $k_-=\sqrt2\kappa=\sqrt2$.}
\e
(\lambda_{a},\lambda_{-})=(-p_a/p_-, -\ln(p_-/k_-)),\label{parameter2}
\ee
where $p_-=p^+\equiv (p^0+p^{D-1})/\sqrt 2$. The vector counterpart of (\ref{ssf1}) $L_\mu{}^\nu(\lambda)$, defined via the equation
\e
L_{\rm S}(\lambda)\g^\mu L^{-1}_{\rm S}(\lambda)=L_\nu{}^\mu(\lambda)\g^\nu,\label{lpmmls}
\ee
is therefore given by
\e
L(\lambda)=\exp(\lambda_a\tau^{+a})\exp(\lambda_{-}\tau^{+-}).\label{ssv1}
\ee
Here $\tau^{+a}=\frac{1}{\sqrt2}(\tau^{0a}+\tau^{D-1,a})$ and $\tau^{+-}=\tau^{0,D-1}$. The  matrix elements of $\tau^{\mu\nu}$ are defined as $(\tau^{\mu\nu})_\s{}^\rho=\d^\mu_\s\eta^{\nu\rho}-\d^\nu_\s\eta^{\mu\rho}$ (see (\ref{tau2})). The matrix elements of $L(\lambda)$ can be either read off from (\ref{lpmmls}) or calculated directly using (\ref{ssv1}): In the lightcone coordinate system, they are given by
\e
&&L_a{}^b(\lambda)=\d_a^b,\quad L_a{}^-(\lambda)=\frac{p_a}{k_-},\quad L_a{}^+(\lambda)=0,\nonumber\\
&&L_-{}^b(\lambda)=0,\quad L_-{}^-(\lambda)=\frac{p_-}{k_-},\quad L_-{}^+(\lambda)=0,\label{llightcn}\\
&&L_+{}^b(\lambda)=-\frac{p^b}{p_-},\quad L_+{}^-(\lambda)=\frac{p_+}{k_-},\quad L_+{}^+(\lambda)=\frac{k_-}{p_-}.\nonumber
\ee
It is straightforward to verify that $L(\lambda)$ does bring $k^\mu$ to $p^\mu$.

The Wigner rotation in spinor space is defined as
\e
W_{\rm S}(\Lambda,\lambda)=L^{-1}_{\rm S}(\lambda_\Lambda)\Lambda_{\rm S}L_{\rm S}(\lambda).\label{littlemlss}
\ee
Here $\Lambda_{\rm S}$ is the general Lorentz transformation in spinor space,
and
\e
L^{-1}_{\rm S}(\lambda_\Lambda)&=&\exp(-\lambda_{\Lambda-}\Sigma^{+-})\exp(-\lambda_{\Lambda a}\Sigma^{+a})\\
&=&\cosh\frac{\lambda_{\Lambda-}}{2}-e^{-\lambda_{\Lambda-}/2}\lambda_{\Lambda a}\Sigma^{+a}-2\sinh\frac{\lambda_{\Lambda-}}{2}\Sigma^{+-},\nonumber
\ee
where the set of parameters $\lambda_\Lambda$ is defined such that $L(\lambda_\Lambda)$ transforms $k^\mu$ into $\Lambda^\mu{}_\nu p^\nu\equiv(\Lambda p)^\mu$, i.e.,
\e
(\lambda_{\Lambda a},\lambda_{\Lambda-})=\bigg(-\frac{(\Lambda p)_a}{(\Lambda p)_-}, \label{llambda} -\ln\frac{(\Lambda p)_-}{k_-}\bigg).
\ee
(The matrix elements of $L(\lambda_\Lambda)$ are given by (\ref{llightcn2}).)

The general Wigner rotation $W_\nu{}^\mu(\Lambda,\lambda)$ can be read off from the following equation:
\e
W_{\rm S}(\Lambda,\lambda)\g^\mu W^{-1}_{\rm S}(\Lambda,\lambda)=W_\nu{}^\mu(\Lambda,\lambda)\g^\nu,
\ee
where in the light-cone coordinates $\g^\mu=(\g^a,\g^-,\g^+)$.

First of all, it is not difficult to verify that (\ref{lgml}) is obeyed,
\e
W_{\rm S}(\Lambda,\lambda)\g^- W^{-1}_{\rm S}(\Lambda,\lambda)=\g^-.
\ee
The above equation implies that
\e
W_b{}^-(\Lambda,\lambda)=W_+{}^-(\Lambda,\lambda)=0\quad{\rm and}\quad W_-{}^-(\Lambda,\lambda)=1.\label{wpp2}
\ee

Secondly, after a length calculation, one obtains
\e
&&W_{\rm S}(\Lambda,\lambda)\g^a W^{-1}_{\rm S}(\Lambda,\lambda)\nonumber\\&=&[(\Lambda_b{}^a+\lambda^a\Lambda_b{}^+)+
(\Lambda_-{}^a+\lambda^a\Lambda_-{}^+)\lambda_{\Lambda b}]\g^b+e^{\lambda_\Lambda}(\Lambda_-{}^a+\lambda^a\Lambda_-{}^+)\g^-.\label{wgw}
\ee
It can be seen that
\e
W_+{}^a(\Lambda,\lambda)&=&0\nonumber\\
W_b{}^a(\Lambda,\lambda)&=&(\Lambda_b{}^a+\lambda^a\Lambda_b{}^+)+
(\Lambda_-{}^a+\lambda^a\Lambda_-{}^+)\lambda_{\Lambda b}\nonumber\\
&=&-\frac{[\Lambda L(\lambda)]_a{}^-[\Lambda L(\lambda)]_-{}^b}{[\Lambda L(\lambda)]_-{}^-}+[\Lambda L(\lambda)]_a{}^b \label{wba}\\
&=&\frac{1}{p_-(\Lambda p)_-}\bigg((p_-\Lambda_b{}^a-p^a\Lambda_b{}^+)(\Lambda p)_--(p_-\Lambda_-{}^a-p^a\Lambda_-{}^+)(\Lambda p)_b\bigg),\nonumber\\
W_-{}^a(\Lambda,\lambda)&=&e^{\lambda_\Lambda}(\Lambda_-{}^a+\lambda^a\Lambda_-{}^+)=\frac{[\Lambda L(\lambda)]_-{}^a}{[\Lambda L(\lambda)]_-{}^-}.\nonumber
\ee
In calculating Eqs. (\ref{wba}), we have used (\ref{llightcn}) and (\ref{llambda}). (The relation between the standard Lorentz transformation $L(\lambda)$ and the momentum $p^\mu$ is given by (\ref{llightcn}).)

Finally, we consider the following equation
\e
W_{\rm S}(\Lambda,\lambda)\g^+ W^{-1}_{\rm S}(\Lambda,\lambda)=W_\nu{}^+(\Lambda,\lambda)\g^\nu.
\ee
We find that the results are
\e
W_+{}^+(\Lambda,\lambda)&=&1,\nonumber\\
W_-{}^+(\Lambda,\lambda)&=&\frac{[\Lambda L(\lambda)]_-{}^+}{[\Lambda L(\lambda)]_-{}^-},\nonumber\\
W_b{}^+(\Lambda,\lambda)&=&-\frac{[\Lambda L(\lambda)]_-{}^+[\Lambda L(\lambda)]_b{}^-}{[\Lambda L(\lambda)]_-{}^-}+[\Lambda L(\lambda)]_b{}^+,\label{wpp}
\ee
where $L(p)$ is defined by (\ref{llightcn}).

Note that the matrix elements in Eqs. (\ref{wpp}) are \emph{not} independent quantities, in the sense that they can be expressed in terms of the other matrix elements by using the Lorentz transformation
\e
W_\mu{}^\rho W_\nu{}^\s\eta_{\rho\s}=\eta_{\mu\nu}.
\ee
For instance, using $W_b{}^\rho W_-{}^\s\eta_{\rho\s}=\eta_{b-}=0$, we obtain that
\e
W_b{}^+(\Lambda,\lambda)=-W_b{}^a(\Lambda,\lambda)W_-{}^a(\Lambda,\lambda)=-\frac{[\Lambda L(\lambda)]_-{}^+[\Lambda L(\lambda)]_b{}^-}{[\Lambda L(\lambda)]_-{}^-}+[\Lambda L(\lambda)]_b{}^+.\label{wbp}
\ee
which is exactly the same as the last equation of (\ref{wpp}). On the other hand, the elements in (\ref{wpp2}) are either $0$ or $1$, so the only ``non-trivial" elements are $W_+{}^a(\Lambda,\lambda)$ and $W_b{}^a(\Lambda,\lambda)$.

Here is another way to calculate the little group element $W_\mu{}^\nu(\Lambda,\lambda)$. First, one can obtain $L_\mu{}^\nu(\lambda_\Lambda)$ by replacing $p^\mu\rightarrow(\Lambda p)^\mu$ and $\lambda\rightarrow \lambda_\Lambda$ in (\ref{llightcn}),
\e
&&L_a{}^b(\lambda_\Lambda)=\d_a^b,\quad L_a{}^-(\lambda_\Lambda)=\frac{(\Lambda p)_a}{k_-},\quad L_a{}^+(\lambda_\Lambda)=0,\nonumber\\
&&L_-{}^b(\lambda_\Lambda)=0,\quad L_-{}^-(\lambda_\Lambda)=\frac{(\Lambda p)_-}{k_-},\quad L_-{}^+(\lambda_\Lambda)=0,\label{llightcn2}\\
&&L_+{}^b(\lambda_\Lambda)=-\frac{(\Lambda p)^b}{(\Lambda p)_-},\quad L_+{}^-(\lambda_\Lambda)=\frac{(\Lambda p)_+}{k_-},\quad L_+{}^+(\lambda_\Lambda)=\frac{k_-}{(\Lambda p)_-}.\nonumber
\ee

Secondly, using the fundamental conditions $L_\mu{}^\rho(\lambda_\Lambda)L_\mu{}^\s(\lambda_\Lambda)\eta_{\rho\s}=\eta_{\mu\nu}$, it is not difficult to determine the inverse of $L_\mu{}^\nu(\lambda_\Lambda)$,
\e
(L^{-1})_\mu{}^\nu(\lambda_\Lambda)=\eta_{\mu\rho}\eta^{\nu\s}L_\s{}^\rho(\lambda_\Lambda).\label{aazh}
\ee
A straightforward computation gives
\e
&&(L^{-1})_a{}^b(\lambda_\Lambda)=\d_a^b,\quad (L^{-1})_a{}^-(\lambda_\Lambda)=-\frac{(\Lambda p)_a}{(\Lambda p)_-},\quad (L^{-1})_a{}^+(\lambda_\Lambda)=0,\nonumber\\
&&(L^{-1})_-{}^b(\lambda_\Lambda)=0,\quad (L^{-1})_-{}^-(\lambda_\Lambda)=\frac{\kappa_-}{(\Lambda p)_-},\quad (L^{-1})_-{}^+(\lambda_\Lambda)=0,\label{llightcnlmda}\\
&&(L^{-1})_+{}^b(\lambda_\Lambda)=\frac{(\Lambda p)^b}{\kappa_-},\quad (L^{-1})_+{}^-(\lambda_\Lambda)=\frac{(\Lambda p)_+}{\kappa_-},\quad (L^{-1})_+{}^+(\lambda_\Lambda)=\frac{(\Lambda p)_-}{\kappa_-}.\nonumber
\ee

Finally, one can calculate all matrix elements $W_\mu{}^\nu(\Lambda,\lambda)$ by substituting (\ref{llightcn}) and (\ref{llightcnlmda}) into the equation
\e
W(\Lambda,\lambda)=L^{-1}( \lambda_\Lambda)\Lambda L(\lambda).
\ee
For instance, using  (\ref{llightcnlmda}), we find that
\e
&&W_b{}^a(\Lambda,\lambda)\nonumber\\
&=&(L^{-1})_b{}^+(\lambda_\Lambda)[\Lambda L(\lambda)]_+{}^a+(L^{-1})_b{}^-(\lambda_\Lambda)[\Lambda L(\lambda)]_-{}^a+(L^{-1})_b{}^c(\lambda_\Lambda)[\Lambda L(\lambda)]_c{}^a\nonumber\\
&=&0-\frac{(\Lambda p)_b}{(\Lambda p)_-}[\Lambda L(\lambda)]_-{}^a+\d_b{}^c[\Lambda L(\lambda)]_c{}^a
\nonumber\\
&=&-\frac{[\Lambda L(\lambda)]_a{}^-[\Lambda L(\lambda)]_-{}^b}{[\Lambda L(\lambda)]_-{}^-}+[\Lambda L(\lambda)]_a{}^b,
\ee
which is exactly the same as the second equation of (\ref{wba}). In the last line, we have used (\ref{llightcn}).

By a length but direct calculation, one can show that
\e
W_a{}^c(\Lambda,\lambda)W_b{}^c(\Lambda,\lambda)=\d_{ab}.\label{soDm2}
\ee
(For a  detailed proof, see Appendix \ref{SecSODm2}.) So $W_b{}^a(\Lambda,\lambda)$ must be the elements of the $SO(D-2)$ subgroup. Hence the group elements $W_b{}^a(\Lambda,\lambda)$ are the most important result of this section. Eq. (\ref{soDm2}) also follows from
\e
\eta_{\mu\nu}W_a{}^\mu(\Lambda,\lambda)W_b{}^\nu(\Lambda,\lambda)=\d_{ab}
\ee
and $W_b{}^-(\Lambda,\lambda)=0$ (see (\ref{wpp2})).

However, we still need to show that the little group is $ISO(D-2)$. Using (\ref{wgw}) and $(\g^-)^2=0$, one obtains immediately
\e
W_{\rm S}(\Lambda,\lambda)
A^a W^{-1}_{\rm S}(\Lambda,\lambda)=W_b{}^a(\Lambda,\lambda)A^b,\label{waw}
\ee
where $A^a=\Sigma^{-a}$ (see (\ref{defA})). On the other hand,
\e
&&W_{\rm S}(\Lambda,\lambda)
\Sigma^{ab} W^{-1}_{\rm S}(\Lambda,\lambda)\nonumber\\
&=&W_c{}^a(\Lambda,\lambda)W_d{}^b(\Lambda,\lambda)\Sigma^{cd}
+(W_-{}^a(\Lambda,\lambda)W_c{}^b(\Lambda,\lambda)-W_-{}^b(\Lambda,\lambda)W_c{}^a(\Lambda,\lambda))A^c. \label{sgmab}
\ee
After defining
\e
a^a(\Lambda,\lambda)\equiv W_-{}^b(\Lambda,\lambda)W_a{}^b(\Lambda,\lambda)=-W_a{}^+(\Lambda,\lambda),\label{trans}
\ee
(See (\ref{wbp}).) Eq. (\ref{sgmab}) can be written as
\e
W_{\rm S}(\Lambda,\lambda)
\Sigma^{ab} W^{-1}_{\rm S}(\Lambda,\lambda)=W_c{}^a(\Lambda,\lambda)W_d{}^b(\Lambda,\lambda)\bigg(\Sigma^{cd}+a^c(\Lambda,\lambda)A^d
-a^d(\Lambda,\lambda)A^c\bigg). \label{sgmab2}
\ee
Eqs. (\ref{waw}) and (\ref{sgmab2}) are the standard transformation law of the set of generators
of $ISO(D-2)$, with the spinor group parameterized as
\e
W_{\rm S}(\Lambda,\lambda)=\exp\bigg(a^a(\Lambda,\lambda)A^a\bigg)\exp\bigg(\frac{1}{2}\Theta_{cd}(\Lambda,\lambda)\Sigma^{cd}\bigg).\label{asigma}
\ee
Here the set of parameters $\Theta_{cd}(\Lambda,\lambda)$ is defined via the equation
\e
\exp\bigg(\frac{1}{2}\Theta_{cd}(\Lambda,\lambda)\tau^{cd}\bigg)_{a}{}^b=W_a{}^b(\Lambda,\lambda),\label{exp}
\ee
with $(\tau^{cd})_a{}^b=\d^c_a\d^{db}-\d^d_a\d^{cb}$.

It is interesting to note that in our construction, the spinor representation matrices of the translation operators $A^a$ satisfy
\e
(A^a)^2=0,\quad\rm (no\ sum)
\ee
where we have used (\ref{defA}). So the eigenvalues of $A^a$ are \emph{zero} automatically, without even considering the topology of the Lorentz group \cite{Weinberg1}. 

Eq. (\ref{asigma}) suggests that the general representation of the little group takes the form
\e
W_{(R)}(\Lambda,\lambda)=\exp\bigg(a^a(\Lambda,\lambda)T_{(R)}^a\bigg)\exp\bigg(
\frac{1}{2}\Theta_{cd}(\Lambda,\lambda)J_{(R)}^{cd}\bigg)\label{wmassless}
\ee
with $T_{(R)}^a$ and $J_{(R)}^{cd}$ furnishing a representation $R$ of the generators of the $ISO(D-2)$ group. However, to avoid continuous degree of freedom of massless particles, we  require that the physical states are eigenstates of $T_{(R)}^a$, but all eigenvalues are zero \cite{Weinberg1}.

\subsection{4 Dimensions, and Applications to Gauge Theory}\label{Seclittgauge}
In $4D$, it is relatively easier to determine the angle of Wigner rotation $\Theta(\Lambda,\lambda)$,
\e
&&\sin(\Theta(\Lambda,\lambda))=W_1{}^2(\Lambda,\lambda)=-W_2{}^1(\Lambda,\lambda),\nonumber\\
&&\cos(\Theta(\Lambda,\lambda))=W_1{}^1(\Lambda,\lambda)=W_2{}^2(\Lambda,\lambda),
\ee
where the matrix elements $W_b{}^a(\Lambda,\lambda)$ ($a, b=1,2$) are given by the second equation of (\ref{wba}).
According to Eq. (\ref{trans}), the set of parameters of the translation part of $ISO(2)$ is
\e
a^a(\Lambda,p)=- W_a{}^+(\Lambda,\lambda),
\ee
whose values can be read off from (\ref{wbp}) and (\ref{llightcn2}).

It is interesting to consider a different  ``standard Lorentz transformation". For instance, let us try
\e
\widetilde L(p)= \exp(-\phi\tau^{12})\exp(-\theta\tau^{13})\exp(\lambda\tau^{03}),\label{Lw}
\ee
with the parameters relating to the momentum $\vec p$ as follows
\e
\hat p^i&=&(\sin\theta\cos\phi,\sin\theta\sin\phi,\cos\theta),\nonumber\\
|\vec p|&=&\kappa e^{-\lambda}.\label{parameterwbg}
\ee
This $\widetilde L(p)$ is adopted from the textbook \cite{Weinberg1}, but rewritten in terms of our notation. It can be seen that $\widetilde L(p)^\mu{}_\nu k^\nu=L(p)^\mu{}_\nu k^\nu=p^\mu$ but $\widetilde L(p)\neq L(p)$. (Our $L(p)$ is defined by (\ref{ssv1}) and (\ref{parameter2}).)
Now the ``new" little group reads
\e
\widetilde W(\Lambda,p)=\widetilde L^{-1}(\Lambda p)\Lambda \widetilde L(p).\label{elements2}
\ee
According to Eq. (\ref{2wigner1}), we must have
\e
\widetilde{W}(\Lambda,p)=S(\Lambda p)W(\Lambda,p)S^{-1}(p).\label{3little}
\ee
Note that
\e
S(p)= \widetilde L^{-1}(p)L(p)\label{s}
\ee
is itself a little group, since
\e
S^\mu{}_\nu(p)k^\nu=(\widetilde L^{-1})^\mu{}_\rho(p)L^\rho{}_\nu(p)k^\nu=(\widetilde L^{-1})^\mu{}_\rho(p)p^\rho=k^\mu.
\ee

In light-cone coordinates, we can decompose Eq. (\ref{3little}) into the following two essential parts
\e
\widetilde{W}_b{}^a(\Lambda,p)&=&S_b{}^c(\Lambda p)W_c{}^d(\Lambda,p)(S^{-1})_d{}^a(p),\label{rotation}\\
\widetilde a^a(\Lambda,p)&=&S_a{}^b(\Lambda p)a^{b}(\Lambda,p)-S_a{}^+(\Lambda p)-S_a{}^b(\Lambda p)W_b{}^c(\Lambda,p)(S^{-1})_c{}^+(p).\label{trans2}
\ee
In deriving (\ref{trans2}), we have used the definition $\widetilde a^a(\Lambda,p)=-\widetilde W_a{}^+(\Lambda,p)$. Eqs (\ref{rotation}) and (\ref{trans2}) also hold in $D$-dimensions.

We now would like to work out $\widetilde W_b{}^a(\Lambda,p)$ ($a, b=1,2$). Inserting (\ref{parameterwbg}) into (\ref{Lw}),
 a direct calculation gives $\wl^\mu{}_\nu(p)$: (We set $\kappa=1$.)
\e
&&\widetilde L^i{}_0(p)=\frac{p_0^2-1}{2p_0^2}p^i,\quad \widetilde L^0{}_0(p)=\frac{p_0^2+1}{2p^0},\quad \widetilde L^i{}_3(p)=\frac{p_0^2+1}{2p_0^2}p^i,\nonumber\\
&&\widetilde L^0{}_3(p)=\frac{p_0^2-1}{2p^0},\quad \widetilde L^a{}_1(p)=\frac{p_3p^a}{p^0\sqrt{p_0^2-p_3^2}},\quad \widetilde L^a{}_2(p)=\frac{-\vp_{ab}p^b}{\sqrt{p_0^2-p_3^2}},\nonumber\\
&&\widetilde L^3{}_1(p)=-\sqrt{1-\frac{p_3^2}{p_0^2}},\quad \widetilde L^3{}_2(p)=\wl^0{}_2(p)=\wl^0{}_1(p)=0,\label{wbgmlss}
\ee
where $\vp_{ab}=-\vp_{ba}$ and $\vp_{12}=1$, and $i=1,2,3$. One can obtain $\wl^\mu{}_\nu(\Lambda p)$ from the above equation by simply replacing $p^\mu$ by $(\Lambda p)^\mu$. The inverse transformation matrix $(\wl^{-1})^\mu{}_\nu(p_\Lambda)$ can be calculated by using the equation $(\wl^{-1})^\mu{}_\nu(p_\Lambda)=\eta^{\mu\rho}\eta_{\nu\s}\wl^\s{}_\rho(p_\Lambda)$; Its expression is
\e
&&(\wl^{-1})^0{}_i(p_\Lambda)=-\frac{(p^0_\Lambda)^2-1}{2(p_\Lambda^0)^2}p_\Lambda^i,\quad (\wl^{-1})^0{}_0(p_\Lambda)=\frac{(p^0_\Lambda)^2+1}{2p_\Lambda^0},\nonumber\\
&& (\wl^{-1})^3{}_i(p_\Lambda)=\frac{(p^0_\Lambda)^2+1}{2(p^0_\Lambda)^2}p_\Lambda^i,\quad(\wl^{-1})^3{}_0(p_\Lambda)=-\frac{(p^0_\Lambda)^2-1}{2p_\Lambda^0},\label{lmdwbg}\\
&& (\wl^{-1})^1{}_a(p_\Lambda)=\frac{p_\Lambda^3p_\Lambda^a}{p_\Lambda^0\sqrt{(p^0_\Lambda)^2-(p_\Lambda^3)^2}},\quad (\wl^{-1})^2{}_a(p_\Lambda)=\frac{-\vp_{ab}p_\Lambda^b}{\sqrt{(p_\Lambda^0)^2-(p^3_\Lambda)^2}},\nonumber\\
&&(\wl^{-1})^1{}_3(p_\Lambda)=-\sqrt{1-\frac{(p_\Lambda^3)^2}{(p_\Lambda^0)^2}},\quad (L^{-1})^2{}_3(p_\Lambda)=(\wl^{-1})^2{}_0(p_\Lambda)=(\wl^{-1})^1{}_0(p_\Lambda)=0,\nonumber
\ee
where $p_\Lambda^\mu$ stands for $(\Lambda p)^\mu$.

In terms of matrix elements, the Wigner rotation (\ref{elements2}) reads
\e
\widetilde W^\mu{}_\nu(\Lambda,p)=(\wl^{-1})^\mu{}_\rho(p_\Lambda)\Lambda^\rho{}_{\s}
\wl^\s{}_\nu(p).\label{wbgwgn}
\ee
Substituting (\ref{wbgmlss}) and (\ref{lmdwbg}) into the above equation, we find that
\e
&&\widetilde W^1{}_1(\Lambda,p)\equiv\cos(\widetilde\Theta(\Lambda,p))\nonumber\\
&=&\frac{\hp^3_\Lambda\hp_\Lambda^a[-\Lambda^a{}_3(1-\hp_3^2)+\Lambda^a{}_b\hp^b\hp^3]
-[1-(\hpl^3)^2][-\Lambda^3{}_3(1-\hp_3^2)+\Lambda^3{}_b\hp^b\hp^3]}{\sqrt{[1-(\hpl^3)^2](1-\hp_3^2)}}
\ee
and
\e
\widetilde W^1{}_2(\Lambda,p)\equiv\sin(\widetilde\Theta(\Lambda,p))=\frac{\vp_{ab}\hp_b(\Lambda^3{}_a
-\Lambda^0{}_a\hpl^3)}{\sqrt{[1-(\hpl^3)^2](1-\hp_3^2)}},
\ee
where the unit vector $\hp^i=p^i/|\vec p|$ is the direction of the momentum $\vec p$, and $\hpl^i$ has a similar definition. Since $\ww^a{}_b(\Lambda,p)$ is an $SO(2)$ matrix, we have $\ww^2{}_2(\Lambda,p)=\ww^1{}_1(\Lambda,p)$ and $\ww^2{}_1(\Lambda,p)=-\ww^1{}_2(\Lambda,p)$.

Similarly, using (\ref{wbgmlss}), (\ref{lmdwbg}), and (\ref{wbgwgn}), the translation part of $ISO(2)$
\e
\widetilde a^a(\Lambda,p)=-\widetilde W_a{}^+(\Lambda,p)
\ee
(see (\ref{trans})) can be worked out, as well. However, since we do not need the explicit expression for $\widetilde a^a(\Lambda,p)$, we do not present it here.

It is interesting to verify (\ref{rotation}) and (\ref{trans2}). One can calculate $S(p)=\widetilde L^{-1}(p)L(p)$ using the definition of $L(p)$ (\ref{llightcn}) and $(\wl^{-1})^\mu{}_\nu(p)=\eta^{\mu\rho}\eta_{\nu\s}\widetilde L^\s{}_\rho(p)$, with $\wl^\s{}_\rho(p)$ defined by (\ref{wbgmlss}). And $S^{-1}(\Lambda p)=L^{-1}(\Lambda p)\wl(\Lambda p)$ can be calculated in a similar way. We have verified (\ref{rotation}) and (\ref{trans2}) in the case of infinitesimal Lorentz transformation
\e
\Lambda^\mu{}_\nu=\d^\mu{}_\nu+(\d\omega)^\mu{}_\nu,
\ee
under the condition that $(p^0)^2-(p^3)^2\neq0$.

We now apply our results to the $U(1)$ gauge theory in $4D$.
In the interaction picture, the  gauge field in $4D$ takes the form \cite{Weinberg1}
\e
a_\mu(x)=\frac{1}{(2\pi)^\frac{3}{2}}\int\frac{d^3p}{\sqrt{2p_0}}\sum_{\s=\pm1}
\bigg[e_{\mu}(\vec p,\s)e^{ip\cdot x}a(\vec p,\s)+e_{\mu}^*(\vec p,\s)e^{-ip\cdot x}a^{\dag}(\vec p,\s)\bigg].
\ee
Here the polarization vector $e^{\mu}(\vec p,\s)=L(p)^\mu{}_\nu e^\mu(\vec k,\s)$, with the standard Lorentz transformation $L(p)^\mu{}_\nu$ defined by (\ref{llightcn}).
Following the convention of \cite{Weinberg1}, we specify the polarization vectors as
\e
e^\mu(\vec k,\pm1)=(1,\pm i,0,0)/\sqrt 2,\nonumber
\ee
where $\vec k$ is the standard momentum.

In $4D$, the vector representation of Eq. (\ref{wmassless}) reads
\e
W^\mu{}_\nu(\Lambda,p)=\exp(a^a(\Lambda,p)\tau^{-a})^\mu{}_\rho
\exp(\Theta(\Lambda,p)\tau^3)^\rho{}_\nu\label{4little}
\ee
where $(\tau^{-a})^\mu{}_\nu=\frac{1}{\sqrt2}(-\tau^{0a}+\tau^{3a})^\mu{}_\nu$, $(\tau^3)^\mu{}_\nu=(\tau^{12})^\mu{}_\nu$, and $(\tau^{\rho\s})^\mu{}_\nu=\eta^{\rho\mu}\d^\s_\nu-
\eta^{\s\mu}\d^\rho_\nu$. From now on, the letter $a$ will be reserved for the creation and annihilation operators, and following the convention of \cite{Weinberg1}, we will denote the translation parameters of $ISO(2)$ as $\a$ and $\b$, namely,
\e
a^a(\Lambda,p)=\bigg(\a(\Lambda,p), \b(\Lambda,p)\bigg).
\ee

Under an arbitrary Lorentz transformation $\Lambda$, the creation and annihilation operators transform as \cite{Weinberg1}
\e
U(\Lambda)a(\vec p,\s)U^{-1}(\Lambda)&=&\sqrt{\frac{(\Lambda p)^0}{p^0}}e^{-i\s\Theta(\Lambda,p)}a(\vec p_\Lambda,\sigma)\label{creation}\\
U(\Lambda)a^\dag(\vec p,\s)U^{-1}(\Lambda)&=&\sqrt{\frac{(\Lambda p)^0}{p^0}}e^{ i\s\Theta(\Lambda,p)}a^\dag(\vec p_\Lambda,\sigma)\label{annihilation}
\ee
Here $\vec p_\Lambda$ stands for $\Lambda^i{}_\mu p^\mu$ or $(\Lambda p)^i$. On the other hand, under
the Lorentz transformation $\Lambda$,
\e
\Lambda^\mu{}_\nu e^\nu(\vec p,\pm1)&=&L^\mu{}_\nu(\Lambda p)(L^{-1}(\Lambda p)\Lambda L(p))^\nu{}_\rho e^\rho(\vec k,\pm1)\nonumber\\
&=&L^\mu{}_\nu(\Lambda p)W^\nu{}_\rho(\Lambda,p) e^\rho(\vec k,\pm1)\nonumber\\
&=&e^{\pm i\Theta(\Lambda,p)}\bigg(e^\mu(\vec p_\Lambda,\pm1)+\frac{\a(\Lambda,p)\pm\b(\Lambda,p)}{|\vec k|}(\Lambda p)^\mu\bigg)
\ee
In the last line, we have used (\ref{4little}).
That is, the polarization vectors cannot transform as a true Lorentz vector \cite{Weinberg1},
\e
e^{-(\pm i\Theta(\Lambda,p))}e_\mu(\vec p,\pm1)=\Lambda^\nu{}_\mu e_\nu(\vec p_\Lambda,\pm1)+\frac{\a(\Lambda,p)\pm i\b(\Lambda,p)}{|\vec k|}p_\mu.\label{iso2}
\ee
Or, according to Weinberg's notation \cite{Weinberg1},
\e
e^\mu(\vec p_\Lambda,\pm1)e^{\pm i\Theta(\Lambda,p)}=\Lambda^\mu{}_\nu e^\nu(\vec p,\pm1)+(\Lambda p)^\mu\Omega_\pm(\Lambda,p)
\ee
Here $\Omega_\pm(\Lambda,p)\equiv -e^{\pm i\Theta(\Lambda,p)}[\a(\Lambda,p)\pm i\b(\Lambda,p)]/|\vec k|$.

So under the Lorentz transformation,
\e
U(\Lambda)a_\mu(x)U^{-1}(\Lambda)=\Lambda^\nu{}_\mu a_\nu(\Lambda x)+\partial_\mu\Omega(x,\Lambda),\label{gauge1}
\ee
where
\e
\Omega(x,\Lambda)=-\frac{i}{(2\pi)^\frac{3}{2}}\int\frac{d^3p}{\sqrt{2p_0}}\sum_{\s=\pm1}
\bigg[\frac{\a+i\b}{|\vec k|}e^{ip\cdot(\Lambda x)}a(\vec p,\s)-\frac{\a-i\b}{|\vec k|}e^{-ip\cdot(\Lambda x)}a^{\dag}(\vec p,\s)\bigg]\label{gaugepara}
\ee
If we calculate everything  using
\e
\widetilde W^\mu{}_\nu(\Lambda,p)=\exp(\widetilde a^a(\Lambda,p)\tau^{-a})^\mu{}_\rho
\exp(\widetilde\Theta(\Lambda,p)\tau^3)^\rho{}_\nu,\label{4little2}
\ee
where
\e
\widetilde a^a(\Lambda,p)=\bigg(\widetilde\a(\Lambda,p), \widetilde\b(\Lambda,p)\bigg),
\ee
in stead of  $W(\Lambda,p)$ (see (\ref{4little})),
the angle $\Theta$ in (\ref{creation}) and (\ref{annihilation}) must be replaced by $\widetilde\Theta$, and $\a$ and $\b$ in (\ref{gaugepara}) must be replaced by $\widetilde\a$ and $\widetilde\b$. (One can transform the set of parameters $(\a,\b,\Theta)$ into $(\widetilde\a,\widetilde\b,\widetilde\Theta)$ by using (\ref{rotation}) and (\ref{trans2}).) After making these replacements, \emph{the only change in} (\ref{gauge1}) is that  $\Omega(x,\Lambda)$  gets replaced by
\e
\widetilde\Omega(x,\Lambda)=-\frac{i}{(2\pi)^\frac{3}{2}}\int\frac{d^3p}{\sqrt{2p_0}}\sum_{\s=\pm1}
\bigg[\frac{\widetilde\a+i\widetilde\b}{|\vec k|}e^{ip\cdot(\Lambda x)}a(\vec p,\s)-\frac{\widetilde\a-i\widetilde\b}{|\vec k|}e^{-ip\cdot(\Lambda x)}a^{\dag}(\vec p,\s)\bigg].
\ee
namely,
\e
U(\Lambda)a_\mu(x)U^{-1}(\Lambda)=\Lambda^\nu{}_\mu a_\nu(\Lambda x)+\partial_\mu\widetilde\Omega(x,\Lambda).\label{gauge2}
\ee
This is the result calculated by using Eq. (\ref{4little2}). 
We see that (\ref{gauge1}) and (\ref{gauge2}) are only up to a gauge transformation, which is due to the \emph{difference} between two ``standard Lorentz transformation", defined  by (\ref{s}). Or in other words, two different ``standard Lorentz transformations" can generate a gauge transformation.

\subsection{Summary of This Section}
In $D$ dimensions, the vector representation of the $SO(D-2)$ part of the Wigner little group $ISO(D-2)$ is given by
\e
W_b{}^a(\Lambda,\lambda)
&=&-\frac{[\Lambda L(\lambda)]_a{}^-[\Lambda L(\lambda)]_-{}^b}{[\Lambda L(\lambda)]_-{}^-}+[\Lambda L(\lambda)]_a{}^b\nonumber\\
&=&\frac{1}{p_-(\Lambda p)_-}\bigg((p_-\Lambda_b{}^a-p^a\Lambda_b{}^+)(\Lambda p)_--(p_-\Lambda_-{}^a-p^a\Lambda_-{}^+)(\Lambda p)_b\bigg),
\ee
and the translation part is defined as
\e
a^a(\Lambda,p)=- W_a{}^+(\Lambda,\lambda)&=&\frac{[\Lambda L(\lambda)]_-{}^+[\Lambda L(\lambda)]_a{}^-}{[\Lambda L(\lambda)]_-{}^-}-[\Lambda L(\lambda)]_a{}^+\nonumber\\
&=&\sqrt2\kappa\bigg(\frac{\Lambda_-{}^+(\Lambda p)^a}{(\Lambda p)_-p_-}-\frac{\Lambda_a{}^+}{p_-}\bigg).
\ee
Here $\Lambda_\mu{}^\nu$ is an arbitrary Lorentz transformation, and  the ``standard Lorentz transformation" $L(\lambda)$
carries the standard $D$-momentum $k^\mu=(0,\ldots,0,\kappa,\kappa)$ to $p^\mu$, i.e. $L^\mu{}_\nu(\lambda)k^\nu=p^\mu$, with $p^\mu$ the $D$-momentum of any massless particle. The matrix $L_\mu{}^\nu(\lambda)$ is defined by (\ref{llightcn}).

The general representation of the little group for massless particles is given by (\ref{wmassless}), where the parameters $\Theta_{cd}$ and $a^a$ defined by (\ref{exp}) and (\ref{trans}), respectively.

\section{Acknowledgement}
This work is supported in part by the National Science Foundation
of China (NSFC) under Grant No. 11475016, and supported partially by the Ren-Cai Foundation of Beijing Jiaotong University through Grant No. 2013RC029, and supported partially by the Scientific Research Foundation for Returned Scholars, Ministry of Education of China.

\appendix

\section{Conventions and Useful Identities}\label{conventions}
In this appendix, we introduce our conventions for the gamma matrices and Clifford algebra of $SO(D-1,1)$, and Lorentz transformations. The set of gamma matrices satisfy
\e
\{\gamma_\mu,\g_\nu\}=2\eta_{\mu\nu},
\ee
where $\eta_{00}=-1$ and $\eta_{ij}=\d_{ij}$. We will use $\eta^{\mu\nu}$ ($\eta_{\mu\nu}$) to raise (lower) indices; For instance, $\g^\mu=\eta^{\mu\nu}\g_\nu$. The gamma matrices obey the reality conditions
\e
\g^{0\dag}=-\g^0,\quad\g^{i\dag}=\g^i.
\ee
The set of generators of $SO(D-1,1)$ are defined as
\e
\Sigma^{\mu\nu}&=&\frac{1}{4}[\g^\mu,\g^\nu].
\ee
It is convenient to decompose the generators into the two sets,
\e
\Sigma^{i0}&=&\frac{1}{4}[\g^i,\g^0]\\
\Sigma^{ij}&=&\frac{1}{4}[\g^i,\g^j]
\ee
They obey the reality conditions
\e
&&\Sigma^{i0\dag}=-\gamma^0\Sigma^{i0}(\gamma^0)^{-1}=\Sigma^{i0},\nonumber\\
&&\Sigma^{ij^\dag}=-\gamma^0\Sigma^{ij}(\gamma^0)^{-1}=-\Sigma^{ij},\label{reality1}
\ee
and satisfy the commutation relations
\e
&&[\Sigma^{\mu\nu},\g^\rho]=\eta^{\nu\rho}\g^\mu-\eta^{\mu\rho}\g^\nu\equiv (\tau^{\mu\nu})_\s{}^\rho\g^\s,\label{tau}\\
&&[\Sigma^{\mu\nu},\Sigma^{\rho\s}]=\eta^{\nu\rho}\Sigma^{\mu\s}-\eta^{\mu\rho}\Sigma^{\nu\s}-\eta^{\nu\s}\Sigma^{\mu\rho}+\eta^{\mu\s}\Sigma^{\nu\rho},
\\
&&\{\Sigma^{\mu\nu},\Sigma^{\rho\s}\}=\frac{1}{2}(\g^{\mu\nu\rho\s}+\eta^{\nu\rho}\eta^{\mu\s}
-\eta^{\mu\rho}\eta^{\nu\s}),\nonumber
\ee
where $\g^{\mu\nu\rho\s}\equiv\g^{[\mu}\g^\nu\g^\rho\g^{\s]}=\frac{1}{4!}(\g^{\mu}\g^\nu\g^\rho\g^{\s}+{\rm permutations})$, and
\e
(\tau^{\mu\nu})_\s{}^\rho=\d^\mu_\s\eta^{\nu\rho}-\d^\nu_\s\eta^{\mu\rho}.\label{tau2}
\ee

We parameterize the general Lorentz transformation $\lds$ in spinor space  as follows
\e
\lds=\exp({\frac{1}{2}\omega_{\mu\nu}\Sigma^{\mu\nu}}),
\ee
where the set of parameters $\omega_{\mu\nu}$ is a real antisymmetric tensor, and the subscript ``S" stands for spinor representation.
Eqs. (\ref{reality1}) imply that $\lds$ obeys the pseudo-reality condition
\e
\g^0\lds^\dag (\g^0)^{-1}=\lds^{-1}.\label{reality2}
\ee
The rotation and boost are given by
\e
R_{\rm S}=e^{\frac{1}{2}\omega_{ij}\Sigma^{ij}}\quad{\rm and}\quad L_{\rm S}=e^{\omega_{i0}\Sigma^{i0}},\label{rnb}
\ee
where $\omega_{i0}$ is the set of rapidities. 

To describe massless particles, it is more convenient to introduce the light-cone coordinates in $D$-dimensional spacetime
\e
x^{\pm}=\frac{1}{\sqrt 2}(\pm x^0+x^{D-1})
\ee
and the transverse space-like coordinates $x^a$, $a=1, 2,\ldots, D-2$.

In terms of light-cone coordinates, we have
\e
\g^{\pm}=\frac{1}{\sqrt 2}(\pm\g^0+\g^{D-1}),
\ee
and the non-vanishing anti-commutators are given by
\e
&&\{\g^+,\g^-\}=2\eta^{+-}=2,\ \nonumber\\
&& \{\g^a,\g^b\}=2\eta^{ab}.
\ee
Hence the metric tensor $\eta^{\mu\nu}$ can be decomposed into
\e
&\eta^{+-}=\eta^{-+}=1,\quad \eta^{ab}=\d^{ab},\quad{\rm and} \quad\eta^{++}=\eta^{--}=\eta^{a+}=\eta^{b-}=0.
\ee
We will use $\eta^{+-}$ or $\eta_{+-}$ to raise or lower indices; For instance, $V_-=\eta_{-+}V^+=V^+$. The inner product of two vectors reads
\e
\eta^{\mu\nu}V_\mu W_\nu=V^aW^a+V_-V^-+V_+V^+.
\ee

Using the rules of tensor analysis, one can write down the general Lorentz transformation $\Lambda$ in the light-cone coordinates;  For instance,
\e
\Lambda_-{}^+=\frac{\partial x^\mu}{\partial x^-}\frac{\partial x^+}{\partial x^\nu}\Lambda_\mu{}^\nu=\frac{1}{2}\bigg(-\Lambda_0{}^0-\Lambda_0{}^{D-1}
+\Lambda_{D-1}{}^0+\Lambda_{D-1}{}^{D-1}\bigg).
\ee
The set of generators $\Sigma^{\mu\nu}$ is decomposed into
\e
&&A^a\equiv\Sigma^{-a}=\frac{1}{4}[\g^-,\g^a],\label{defA}\\
&&\Sigma^{+-}=\frac{1}{4}[\gamma^+,\gamma^-]=\Sigma^{0,D-1},\\
&&\Sigma^{+a}=\frac{1}{4}[\g^+,\g^a],
\\
&&\Sigma^{ab}=\frac{1}{4}[\g^a,\g^b].
\ee
Under the above decomposition, the (spinor) algebra of the little group $ISO(D-2)$ reads
\e
&&[A^a, A^b]=0,\\
&&[\Sigma^{ab},A^c]=\d^{bc}A^a-\d^{ac}A^b,\\
&&[\Sigma^{ab},\Sigma^{cd}]=\d^{bc}\Sigma^{ad}-\d^{ac}\Sigma^{bd}-\d^{bd}\Sigma^{ac}+\d^{ad}\Sigma^{bc}.
\ee
Notice that by the definition of $A^a$ (see (\ref{defA})),
\e
(A^a)^2=0,
\ee
that is, in the spinor representation, the eigenvalues of $A^a$ are zero automatically.
\section{Verifying Little Group $SO(D-1)$}\label{SecSODm1}
We now try to give a direct verification of (\ref{soDm1}), which is essentially the same as the following equation:
\e
W_i{}^k(\Lambda,p)W_j{}^k(\Lambda,p)=\d_{ij} .\label{soDm12}
\ee
For readability, we will write $[\Lambda L(p)]_\mu{}^\nu$ as $(\Lambda L)_\mu{}^\nu$. Our main equation for proving (\ref{soDm12}) is the fundamental one:
\e
\eta_{\rho\sigma}(\Lambda L)_\mu{}^\rho(\Lambda L)_\nu{}^\sigma=\eta_{\mu\nu}\quad{\rm or}\quad (\Lambda L)_\mu{}^k(\Lambda L)_\nu{}^k=\eta_{\mu\nu}+(\Lambda L)_\mu{}^0(\Lambda L)_\nu{}^0.\label{fdmtl}
\ee
Inserting the last equation of (\ref{gwm1}) into the left-hand side of (\ref{soDm12}) gives
\e
&&W_i{}^k(\Lambda,p)W_j{}^k(\Lambda,p)\nonumber\\
&=&\frac{1}{[1+(\Lambda L)_0{}^0]^2}\bigg((\Lambda L)_0{}^k(\Lambda L)_0{}^k(\Lambda L)_i{}^0(\Lambda L)_j{}^0-(\Lambda L)_i{}^k(\Lambda L)_0{}^k(\Lambda L)_0{}^0(\Lambda L)_j{}^0\nonumber\\
&&-(\Lambda L)_i{}^k(\Lambda L)_0{}^k(\Lambda L)_j{}^0-(\Lambda L)_0{}^k(\Lambda L)_j{}^k(\Lambda L)_i{}^0(\Lambda L)_0{}^0+(\Lambda L)_i{}^k(\Lambda L)_j{}^k(\Lambda L)_0{}^0(\Lambda L)_0{}^0\nonumber\\
&&+(\Lambda L)_i{}^k(\Lambda L)_j{}^k(\Lambda L)_0{}^0-(\Lambda L)_0{}^k(\Lambda L)_j{}^k(\Lambda L)_i{}^0+(\Lambda L)_i{}^k(\Lambda L)_j{}^k(\Lambda L)_0{}^0+(\Lambda L)_i{}^k(\Lambda L)_j{}^k\bigg)\nonumber\\ \label{ap1}
\ee
The summation of the first term of first line in the big bracket of (\ref{ap1}) and the third term of the second line is
\e
&&\bigg((\Lambda L)_0{}^k(\Lambda L)_0{}^k\bigg)(\Lambda L)_i{}^0(\Lambda L)_j{}^0+\bigg((\Lambda L)_i{}^k(\Lambda L)_j{}^k\bigg)(\Lambda L)_0{}^0(\Lambda L)_0{}^0\nonumber\\
&=&\bigg(\eta_{00}+(\Lambda L)_0{}^0(\Lambda L)_0{}^0\bigg)(\Lambda L)_i{}^0(\Lambda L)_j{}^0+\bigg(\d_{ij}+(\Lambda L)_i{}^0(\Lambda L)_j{}^0\bigg)(\Lambda L)_0{}^0(\Lambda L)_0{}^0\nonumber\\
&=&[\d_{ij}+2(\Lambda L)_i{}^0(\Lambda L)_j{}^0][(\Lambda L)_0{}^0]^2-(\Lambda L)_i{}^0(\Lambda L)_j{}^0.\label{ap2}
\ee
Let us now add the second term of first line in the bracket of (\ref{ap1}) and the second term of the second line,
\e
&&-\bigg((\Lambda L)_i{}^k(\Lambda L)_0{}^k\bigg)(\Lambda L)_0{}^0(\Lambda L)_j{}^0-\bigg((\Lambda L)_0{}^k(\Lambda L)_j{}^k\bigg)(\Lambda L)_i{}^0(\Lambda L)_0{}^0\nonumber\\
&=&-\bigg(\eta_{i0}+(\Lambda L)_i{}^0(\Lambda L)_0{}^0\bigg)(\Lambda L)_0{}^0(\Lambda L)_j{}^0-\bigg(\eta_{0j}+(\Lambda L)_0{}^0(\Lambda L)_j{}^0\bigg)(\Lambda L)_i{}^0(\Lambda L)_0{}^0\nonumber\\
&=&-2(\Lambda L)_i{}^0(\Lambda L)_j{}^0[(\Lambda L)_0{}^0]^2.\label{ap3}
\ee
The summation of the rest terms (the first term of second line and all terms of third line) in the big bracket of (\ref{ap1}) is
\e
&&-(\Lambda L)_i{}^k(\Lambda L)_0{}^k(\Lambda L)_j{}^0+(\Lambda L)_i{}^k(\Lambda L)_j{}^k(\Lambda L)_0{}^0-(\Lambda L)_0{}^k(\Lambda L)_j{}^k(\Lambda L)_i{}^0\nonumber\\
&&+(\Lambda L)_i{}^k(\Lambda L)_j{}^k(\Lambda L)_0{}^0+(\Lambda L)_i{}^k(\Lambda L)_j{}^k
\nonumber\\
&=&-\bigg((\Lambda L)_i{}^k(\Lambda L)_0{}^k\bigg)(\Lambda L)_j{}^0-\bigg((\Lambda L)_0{}^k(\Lambda L)_j{}^k\bigg)(\Lambda L)_i{}^0+\bigg((\Lambda L)_i{}^k(\Lambda L)_j{}^k\bigg)[2(\Lambda L)_0{}^0+1]
\nonumber\\
&=&-\bigg(\eta_{i0}+(\Lambda L)_i{}^0(\Lambda L)_0{}^0\bigg)(\Lambda L)_j{}^0-\bigg(\eta_{0j}+(\Lambda L)_0{}^0(\Lambda L)_j{}^0\bigg)(\Lambda L)_i{}^0\nonumber\\
&&+\bigg(\d_{ij}+(\Lambda L)_i{}^0(\Lambda L)_j{}^0\bigg)[2(\Lambda L)_0{}^0+1]
\nonumber\\
&=&-2(\Lambda L)_i{}^0(\Lambda L)_j{}^0(\Lambda L)_0{}^0+\bigg(\d_{ij}+(\Lambda L)_i{}^0(\Lambda L)_j{}^0\bigg)[2(\Lambda L)_0{}^0+1]\label{ap4}
\ee
In deriving (\ref{ap2}), (\ref{ap3}), and (\ref{ap4}), we have used (\ref{fdmtl}). The big bracket of (\ref{ap1}) is the summation of (\ref{ap2}), (\ref{ap3}), and (\ref{ap4}):
\e
{\rm (\ref{ap2})+(\ref{ap3})+(\ref{ap4})}=\d_{ij}\bigg(1+2(\Lambda L)_0{}^0+[(\Lambda L)_0{}^0]^2\bigg).\label{ap5}
\ee
Replacing the big bracket of (\ref{ap1}) by (\ref{ap5}), the right-hand side of (\ref{ap1}) becomes $\d_{ij}$. This completes the proof of (\ref{soDm12}).


\section{Verifying Little Group $SO(D-2)$}\label{SecSODm2}
We now give a direct proof of (\ref{soDm2}). For convenience, we cite it here:
\e
W_a{}^c(\Lambda,p)W_b{}^c(\Lambda,p)=\d_{ab}.\label{app1}
\ee
We are going to use the fundamental equation
\e
\eta^{\rho\sigma}(\Lambda L)_\rho{}^\mu(\Lambda L)_\sigma{}^\nu&=&\eta^{\mu\nu}\nonumber\\
{\rm or}\quad(\Lambda L)_c{}^\mu(\Lambda L)_c{}^\nu&=&\eta^{\mu\nu}-(\Lambda L)_+{}^\mu(\Lambda L)_-{}^\nu-(\Lambda L)_-{}^\mu(\Lambda L)_+{}^\nu\label{fdmt2}
\ee
to prove (\ref{app1}), where we have written  $[\Lambda L(p)]_\mu{}^\nu$ as $(\Lambda L)_\mu{}^\nu$.
Plugging the second line of the second equation of (\ref{wba}) into the left-hand side of (\ref{app1}),
\e
W_a{}^c(\Lambda,p)W_b{}^c(\Lambda,p)&=&\frac{(\Lambda L)_c{}^-(\Lambda L)_c{}^-(\Lambda L)_-{}^a(\Lambda L)_-{}^b}{[(\Lambda L)_-{}^-]^2}\nonumber\\
&&-\frac{(\Lambda L)_c{}^-(\Lambda L)_c{}^b(\Lambda L)_-{}^a+(\Lambda L)_c{}^-(\Lambda L)_c{}^a(\Lambda L)_-{}^b}{(\Lambda L)_-{}^-}\nonumber\\
&&+(\Lambda L)_c{}^a(\Lambda L)_c{}^b.
\label{app2}
\ee
According to (\ref{fdmt2}),
\e
(\Lambda L)_c{}^-(\Lambda L)_c{}^-&=&\eta^{--}-(\Lambda L)_+{}^-(\Lambda L)_-{}^--(\Lambda L)_-{}^-(\Lambda L)_+{}^-\nonumber\\
&=&-2(\Lambda L)_-{}^-(\Lambda L)_+{}^-\label{fdmt3}
\ee
Taking account of (\ref{fdmt3}), the first line of (\ref{app2}) becomes
\e
\frac{1}{[(\Lambda L)_-{}^-]^2}\bigg((\Lambda L)_c{}^-(\Lambda L)_c{}^-\bigg)(\Lambda L)_-{}^a(\Lambda L)_-{}^b=-\frac{2(\Lambda L)_+{}^-(\Lambda L)_-{}^a(\Lambda L)_-{}^b}{(\Lambda L)_-{}^-}.\label{app3}
\ee
Similarly, one can convert the second of (\ref{app2}) into the form:
\e
&&-\frac{1}{(\Lambda L)_-{}^-}\bigg[\bigg((\Lambda L)_c{}^-(\Lambda L)_c{}^b\bigg)(\Lambda L)_-{}^a+\bigg((\Lambda L)_c{}^-(\Lambda L)_c{}^a\bigg)(\Lambda L)_-{}^b\bigg]\nonumber\\
&=&-\frac{1}{(\Lambda L)_-{}^-}\bigg[\bigg(\eta^{-b}-(\Lambda L)_+{}^-(\Lambda L)_-{}^b-(\Lambda L)_-{}^-(\Lambda L)_+{}^b\bigg)(\Lambda L)_-{}^a+(a\leftrightarrow b)\bigg]\nonumber\\
&=&\frac{2(\Lambda L)_+{}^-(\Lambda L)_-{}^a(\Lambda L)_-{}^b}{(\Lambda L)_-{}^-}+
(\Lambda L)_+{}^a(\Lambda L)_-{}^b+
(\Lambda L)_-{}^a(\Lambda L)_+{}^b\label{app4}
\ee
Inserting (\ref{app3}) and (\ref{app4}) into (\ref{app2}),
\e
W_a{}^c(\Lambda,p)W_b{}^c(\Lambda,p)&=&(\Lambda L)_+{}^a(\Lambda L)_-{}^b+
(\Lambda L)_-{}^a(\Lambda L)_+{}^b
+(\Lambda L)_c{}^a(\Lambda L)_c{}^b
\nonumber\\
&=&\d_{ab}.
\ee
This completes the proof.

\end{document}